\documentclass[preprint,12pt]{elsarticle}

\usepackage{amssymb}
\usepackage[mathscr]{eucal}
\usepackage{setspace}
\usepackage{float}
\usepackage{array}
\usepackage{multirow}

\usepackage{amsmath}
\usepackage{geometry}
\usepackage{graphics}
\graphicspath{{Figures/}}

\usepackage{hyperref}
\hypersetup{
  colorlinks   = true, 
  urlcolor     = blue, 
  linkcolor    = blue, 
  citecolor   = blue 
}

\biboptions{sort&compress}
\usepackage{blindtext}

\journal{Journal of Sound and Vibrations}

\begin{document}

\begin{frontmatter}



\title{Abrupt transitions in turbulent thermoacoustic systems}

\author[inst1]{Ramesh S. Bhavi}
\ead{rameshbhavi003@gmail.com}
\author[inst2]{Induja Pavithran}
\author[inst1]{Amitesh Roy}
\author[inst1]{R. I. Sujith}

\affiliation[inst1]{organization={Department of Aerospace Engineering},
            addressline={IIT Madras}, 
            city={Chennai},
            postcode={600036}, 
            country={India}}

\affiliation[inst2]{organization={Department of Physics},
            addressline={IIT Madras}, 
            city={Chennai},
            postcode={600036}, 
            country={India}}

\begin{abstract}

Abrupt transitions to the state of thermoacoustic instability (TAI) in gas turbine combustors are a significant challenge plaguing the development of next-generation low-emission aircraft and power generation engines. In this paper, we present the observation of abrupt transition in three disparate turbulent thermoacoustic systems: an annular combustor, a swirl-stabilized combustor, and a preheated bluff-body stabilized combustor. Using a low-order stochastic thermoacoustic model, we show that the reported abrupt transitions occur when an initially stable, supercritical limit cycle becomes unstable, leading to a secondary bifurcation to a large amplitude limit cycle solution. The states of combustion noise and intermittency observed in these turbulent combustors are well captured by the additive stochastic noise in the model. Through amplitude reduction, we analyze the underlying potential functions affecting the stability of the observed dynamical states. Finally, we make use of the Fokker-Planck equation, educing the effect of stochastic fluctuations on subcritical and secondary bifurcation. We conclude that a high enough intensity of stochastic fluctuations which transforms a subcritical bifurcation into an intermittency-facilitated continuous transition may have little effect on the abrupt nature of secondary bifurcation. Our findings imply the high likelihood of abrupt transitions in turbulent combustors possessing higher-order nonlinearities where turbulence intensities are disproportionate to the large amplitude limit cycle solution.  
\end{abstract}




\end{frontmatter}



\section{Introduction}
Thermoacoustic instability (TAI) manifests as large amplitude pressure oscillations in gas turbine and rocket combustors. These large amplitude oscillations lead to loss of structural integrity through mechanical vibrations and cause the failure of thermal protection systems due to enhanced heat transfer \cite{lieuwen2005combustion}. TAI occurs through a feedback coupling between the heat release rate fluctuations arising from the unsteady flame and the acoustic field of the combustor \cite{juniper2018sensitivity, sujith2021thermoacoustic}. Such positive feedback leads to runaway growth in the amplitude of pressure oscillations. The growth is counterbalanced by increased acoustic losses, which leads to the state of limit cycle oscillations observed during TAI \cite{rayleigh1878explanation, chu1965energy}. 

Limit cycle oscillations emerge due to underlying nonlinearities of thermoacoustic systems. The transition of a dynamical system from a fixed point to a limit cycle solution owing to a change in the control parameter is termed as Hopf bifurcation \cite{strogatz2018nonlinear}. If, during the transition, the amplitude of the limit cycle increases gradually, then it is referred to as a supercritical Hopf bifurcation. If, on the contrary, the transition is abrupt, it is called a subcritical Hopf bifurcation. \citet{lieuwen2002experimental} described the transition from the state of stable combustor operation to the state of TAI as a Hopf bifurcation from a fixed point to a limit cycle solution. Several experimental studies have since then reported supercritical and subcritical bifurcation to the state of limit cycles \cite{moeck2008subcritical,li2017experimental,juniper2012triggering,etikyala2017change,subramanian2010bifurcation}. \citet{ananthkrishnan1998application, ananthkrishnan2005reduced} hypothesized the possibility of a secondary bifurcation from an initially stable primary limit cycle to a large amplitude secondary limit cycle solution in thermoacoustic systems having higher-order nonlinearities. Secondary bifurcation was then experimentally confirmed in laminar \cite{nalininonlinear} and, very recently, in turbulent \cite{roy2021flame, singh2021intermittency, wang2021multi} thermoacoustic systems.

An aspect that is overlooked in approaching the problem of thermoacoustic transitions in gas turbine combustors is that the state of stable combustor operation is seldom a fixed point. This state is better characterised by aperiodic fluctuations arising due to turbulence and is referred to as combustion noise \cite{candel2009flame,gotoda2011dynamic}. In fact, the aperiodic fluctuations during combustion noise have chaotic and multifractal signatures \cite{nair2013loss, tony2015detecting,nair2014multifractality}.  \citet{nair2014intermittency} reported that the change of the state of a system from combustion noise to limit cycle oscillation takes place through the state of intermittency. The state of intermittency is an intermediate state and is characterised by bursts of periodic high amplitude oscillations amidst epochs of aperiodic low amplitude fluctuations. Thus, intermittency has the imprint of both combustion noise and TAI. Transition to TAI through intermittency has been confirmed in many studies since \cite{gotoda2014detection, huang2015advanced, kabiraj2015chaos, kheirkhah2017dynamics}. The occurrence of intermittency leads to a smooth variation of statistical measures of the system, such as the root-mean-squared (rms) or Fourier amplitude, as the state of a system changes from a state of combustion noise to TAI.

In modeling intermittent transitions, the state of combustion noise is often assumed to be of stochastic origin \cite{clavin1994turbulence, burnley2000influence} in view of the difficulty in modeling pressure fluctuations that have chaotic and multifractal characteristics. Thus, modeling studies incorporate the fluctuations as additive \cite{burnley2000influence,noiray2013deterministic,clavin1994turbulence, noiray2017method, bonciolini2017subcritical,gopalakrishnan2016stochastic} and multiplicative noise \cite{kasthuri2019bursting,clavin1994turbulence,burnley2000influence} in models of supercritical and subcritical bifurcation. These stochastic models are then analyzed by deriving the Fokker-Planck equation from which a stationary solution is obtained \cite{gopalakrishnan2016stochastic,bonciolini2017subcritical,noiray2017method}. The solution of the Fokker-Planck equation yields the evolution of the probability density function (PDF) of the envelope of the amplitude of fluctuations during the transition. For instance, \citet{gopalakrishnan2016stochastic} showed that abrupt subcritical bifurcation in a laminar thermoacoustic system becomes continuous at high enough noise intensity. Other approaches have used phenomenological models of vortex shedding \cite{nair2015reduced,seshadri2016reduced} and by incorporating explicit slow and fast time scales in lower-order models \cite{tandon2020bursting}.

The occurrence of both abrupt and continuous transitions in thermoacoustic systems makes apparent the significant challenge in their modeling. In addition, the observation of abrupt secondary transition \cite{roy2021flame, singh2021intermittency, wang2021multi} in highly turbulent thermoacoustic systems is not understood clearly. Specifically, the explanation of what makes a transition continuous and another abrupt has been found lacking in the literature. 

In this paper, we begin by presenting evidence of abrupt secondary bifurcation arising in three different turbulent thermoacoustic systems: annular combustor, swirl-stabilized dump combustor, and bluff-body stabilized dump combustor with preheated air. We use a stochastic thermoacoustic model based on a modified Van der Pol oscillator containing higher-order nonlinearities \cite{ananthkrishnan1998application} to describe the secondary bifurcations to a very high-amplitude TAI. Through the application of deterministic and stochastic averaging, we obtain the evolution equations for the amplitude and the phase of the oscillations. Using the amplitude equation and its potential function, we explain the stability of different dynamical states of the thermoacoustic systems during secondary bifurcation. Finally, we derive and solve the Fokker-Planck equation and explain the effect of stochastic fluctuations on the evolution of the PDF of pressure fluctuations during the transitions.

The rest of the paper is categorized as follows. Section \ref{Experimental setup} describes the setup and methods used for conducting experiments. The observations of abrupt transitions via secondary bifurcations in the experiments are detailed in Sec. \ref{describing Experimental results}. The thermoacoustic model describing secondary bifurcation is presented in Sec. \ref{model to describe secondary bifurcations}. Section \ref{secondary bifurcation of the deterministic system} describes the role of higher-order nonlinear coefficients in obtaining primary and secondary limit cycles. Section \ref{Slow flow representation of the stochastic system} describes the stochastic model from which the slowly varying amplitude and phase equations are derived. The PDF of the amplitude envelope is obtained with the help of the Fokker-Planck equation in Sec. \ref{Stationary solution of the Fokker-Planck equation}. The effect of stochastic fluctuations on the transition is explained in Sec. \ref{Effect of noise levels on the transition}. The stability of different dynamical states is visualized with the help of potential function in Sec. \ref{Potential landscape of the secondary bifurcation}. Section \ref{Conclusion} summarizes the conclusions from the paper.


\section{Experiments}\label{Experimental setup}
We performed experiments in three different turbulent combustors to obtain abrupt transitions to the state of TAI. The combustor setups are shown in Fig. \ref{experimentalsetups} and detailed below.

\subsection{Annular combustor}
\label{Sec:annular_setup}
\begin{figure}[h]
\centering
\includegraphics[width=13cm]{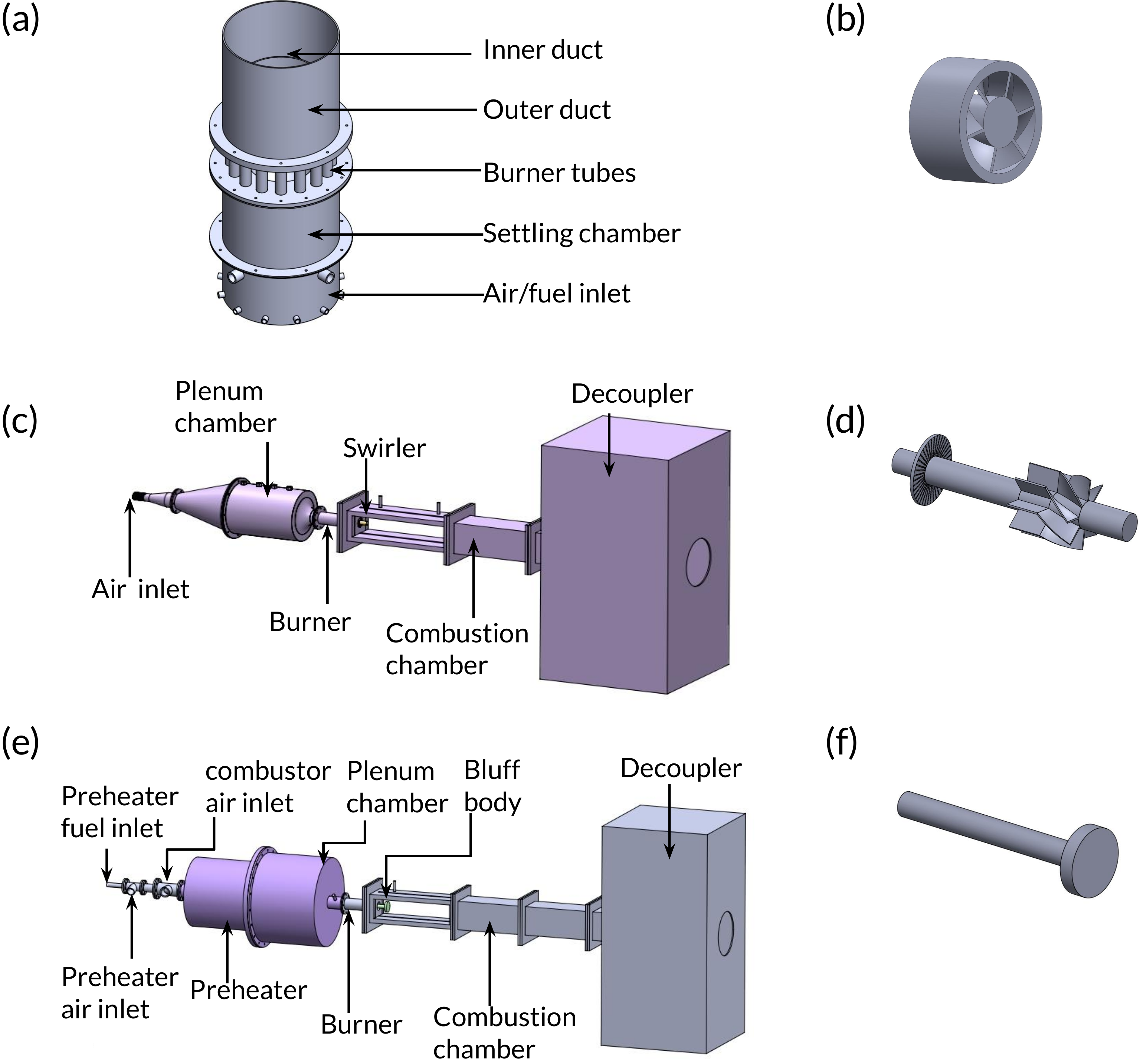}
\caption{Schematic of (a) annular combustor comprising sixteen swirl-stabilized burners, (c) swirl-stabilized dump combustor and (e) bluff-body stabilized dump combustor with preheated air. The flame stabilizing mechanism for the three combustors are: swirler in (b), (d) and bluff-body in (f), respectively.}
    \label{experimentalsetups}
\end{figure}

Figure \ref{experimentalsetups}(a) depicts the swirl-stabilized annular combustor. The air/fuel inlet is connected to the settling chamber containing a flow straightener for reducing flow non-uniformity and a hemispherical flow divider. The settling chamber bolsters 16 burners arranged in an annular arrangement. The diameter of the burner tubes is 30 mm, and their length is 150 mm. These burners are connected to the combustion chamber comprising an inner and outer duct. The length of the outer and inner ducts are 400 mm and 200 mm, respectively. Each burner houses a swirler for flame stabilization. Each swirler possesses six vanes inclined at an angle of $\beta = 60^{\circ}$ with the burner axis (cf. Fig. \ref{experimentalsetups}b).

Partially premixed air and fuel (LPG, 40\% propane, and 60\% butane) are used for the experiments. The combustor is ignited with the help of a non-premixed pilot flame, which is switched off following flame stabilization. The equivalence ratio ($\phi$) is increased from 0.40 to 0.58 in a quasi-static manner by varying the fuel flow rate. The airflow rate is kept constant at 1400 SLPM throughout the experiments. In the forward direction, the fuel flow rate is increased from 40 to 45.6 SLPM, while it is decreased from 48 to 36 SLPM in the reverse direction. The Reynolds number, calculated using the exit diameter of the burner ($d = 15$ mm), varies from $Re_d \approx 0.56 \times 10^4 $ to $1.22 \times 10^4$. The power of the combustor based on the fuel flow rate varies between 39 and 79 kW. Please refer to \cite{ singh2021intermittency,roy2021flame} for more information on the annular combustor setup.

\subsection{Swirl-stabilized combustor}
\label{Sec: SwirlCombustor}
Figure \ref{experimentalsetups}(c) shows the schematic of the swirl-stabilized combustor. The setup consists of a plenum chamber, combustion chamber, and an acoustic decoupler. A central shaft through the burner is used as a fuel pipe to deliver the fuel. Compressed air is passed, in a co-axial manner around the fuel pipe, to the plenum chamber. The air and the fuel are mixed in the burner at a distance of 85 mm from the combustor dump plane. The burner has a diameter of 40 mm. The combustion chamber has a square cross-section ($90 \times 90$ mm$^2$) and is 800 mm in length. The combustor exhausts into the acoustic decoupler, which maintains acoustically open boundary conditions at the combustor exit. A swirler of diameter $d = 40$ mm consisting of 8 vanes, with each vane having an angle of $40^{\circ}$ with respect to the longitudinal axis, is used for flame stabilization (cf. Fig. \ref{experimentalsetups}d). The swirler is mounted such that the end of each vane is flush mounted to the dump plane.   

Partially premixed air and LPG are used for performing the experiments. The equivalence ratio is varied as the control parameter by increasing the airflow rate from 380 SLPM to 700 SLPM in steps of 20 SLPM. The fuel flow rate is maintained constant at 24 SLPM. The airflow rate is varied in a quasi-static manner. Thus, the value of $\phi$ decreases from $0.99$ to $0.54$. The power of the combustor based on the given fuel flow rate is 24.30 kW. The Reynolds number, determined using the diameter of the swirler, varies between of $Re_d = 1.45 \times 10^4 $ and $2.60 \times 10^4$.

\subsection{Preheated bluff-body stabilized combustor}
\label{preheatersetup}
The schematic of the preheated combustor is shown in Fig. \ref{experimentalsetups}(e). The combustor comprises a preheater, plenum chamber, combustion chamber, and an acoustic decoupler. A portion of the airflow is bypassed to the preheater. The hot gases from the preheater are then mixed with the airflow prior to the main combustor at the burner to increase the temperature of the air and fuel. The fuel for the main stage is injected through the central shaft supporting the bluff body into the burner just before the dump plane. A bluff-body (cf. Fig. \ref{experimentalsetups}f) is used for flame stabilization. The diameter of the bluff-body is $d = 47$ mm. The cross-section and length of the combustor are $90 \times 90$ mm$^2$ and 1200 mm, respectively.

Experiments are conducted by varying the equivalence ratio ($\phi$) of the main combustor as the control parameter by keeping the fuel flow rate constant (30 SLPM) and increasing the airflow rate from 800 SLPM to 1350 SLPM in steps of 50 SLPM. The airflow rate is varied in a quasi-static manner. The equivalence ratio varies from $\phi=0.67$ to 0.35. The power of the main combustor based on the given fuel flow rate is 30.3 kW. Based on the diameter of the bluff-body, the Reynolds number varies in the range of $Re_d =3.36 \times 10^4$ to $5.67 \times 10^4$. For the experiments reported here, the preheating temperature was maintained at $T= 300$$^\circ$C. Please refer to \cite{pawar2021effect} for more details on the preheater setup.

\subsection{Instrumentation}
The flow rates of air and fuel in all these experiments are controlled using mass flow controllers (Alicat mass flow controllers, MCR series) with a measurement uncertainty of $\pm$(0.8 $\%$ of reading + 0.2 $\%$ of full scale). The maximum error in the reported value of $\phi$ is $\pm 1.6\%$, and for $Re$ it is $\pm 0.8 \%$. In all experiments, the control parameter ($\phi$) is varied in a quasi-static manner. Piezotronics PCB103B02 make piezoelectric pressure transducers are used for pressure fluctuations measurements. The sensitivity of the transducers is 217.5 mV/kPa. The signal from the pressure transducer is acquired for a duration of 3 s at a sampling rate of 10 kHz for the annular combustor and swirl stabilized dump combustor, and 20 kHz for the bluff-body stabilized dump combustor. The maximum uncertainty in the reported values of pressure measurement is $\pm 0.15$ Pa. Semi-infinite waveguides of length 3.2 m (annular combustor), and 10 m (dump combustors) are attached to the pressure mountings. The inner diameter of the waveguides is 4 mm. Further details on the instrumentation can be found in \cite{roy2021flame, pawar2021effect}.

\begin{figure}
\centering
\includegraphics[width=\textwidth]{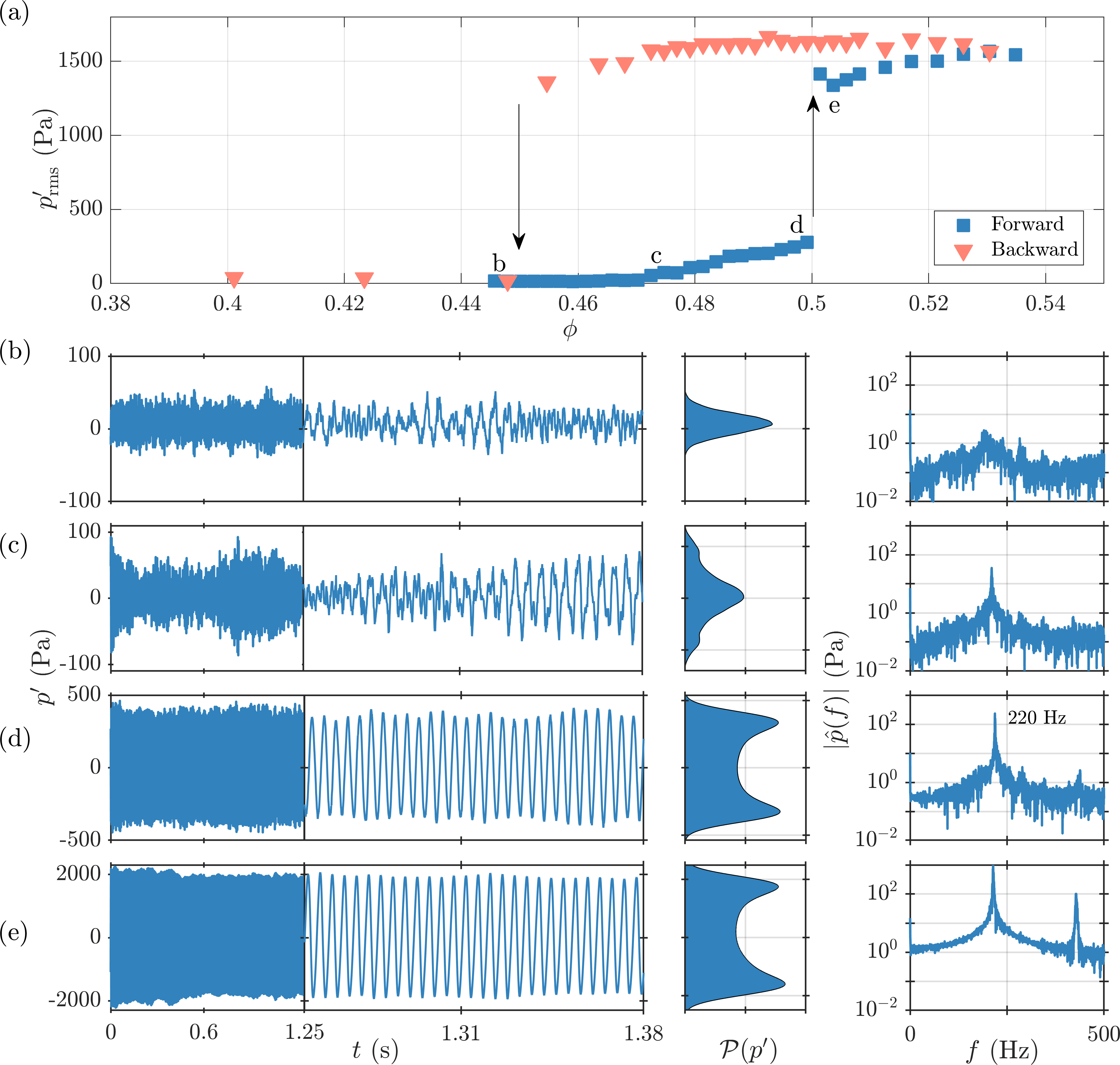}
\caption{Characteristics of secondary bifurcation in the annular combustor. (a) The variation of $p'_\mathrm{rms}$ as a function of the control parameter $\phi$. Panels (b-e) shows the time series, the PDF $\mathcal{P}(p')$ and the amplitude spectrum $|\hat{p}(f)|$ of pressure fluctuations $p'$ observed during the states of (b) combustion noise, (c) intermittency, (d) low amplitude limit cycle oscillations and (e) large amplitude limit cycle oscillations, as indicated in panel (a). Note the increase in the abscissa limits for the time series and distribution in panels (b) to (e).}
\label{Annualr timeseries}
\end{figure}

\section{Secondary bifurcation in turbulent combustors}\label{describing Experimental results}
Let us begin by considering the characteristics of the bifurcation when the equivalence ratio ($\phi$) is changed in the turbulent combustors. Figure \ref{Annualr timeseries}(a) depicts the variation in $p^\prime_\mathrm{rms}$ when the control parameter $\phi$ is increased in the annular combustor. For low values of equivalence ratio ($\phi<0.46$), the state of the system is characterised by combustion noise (cf. Fig. \ref{Annualr timeseries}b) possessing very low amplitude ($p'_{\text{rms}} \approx 20$ Pa) of fluctuations. The fluctuations are characterised by a unimodal distribution and a broadband spectrum. Upon increasing the equivalence ratio ($\phi$) beyond a value of (0.46), we observe the state of intermittency, where aperiodic fluctuations are randomly interspersed with bursts of periodic oscillations (cf. Fig. \ref{Annualr timeseries}c). The appearance of periodic bursts, whose amplitude is higher than the amplitude of combustion noise, alters the initially unimodal distribution; we observe secondary peaks at $|p^\prime|\neq0$ (see PDF in Fig. \ref{Annualr timeseries}c). The increased periodic content appears as a narrowband peak in the amplitude spectrum. Upon further increasing the value of $\phi$, we observe the state of low amplitude limit cycle oscillations (LCO) with $p'_\text{rms} \approx$ 373 Pa. The limit cycle oscillations show (cf. Fig \ref{Annualr timeseries}d) a bi-modal distribution and a narrowband peak in the amplitude spectrum at 220 Hz. Finally, for $\phi>0.50$, we observe (cf. Fig \ref{Annualr timeseries}e) an abrupt transition from the low amplitude primary limit cycle oscillations to a large amplitude ($p'_{\text{rms}} \approx 1500 $ Pa) secondary limit cycle oscillations. 

\begin{figure}[h]
\centering
\includegraphics[width=0.95\textwidth]{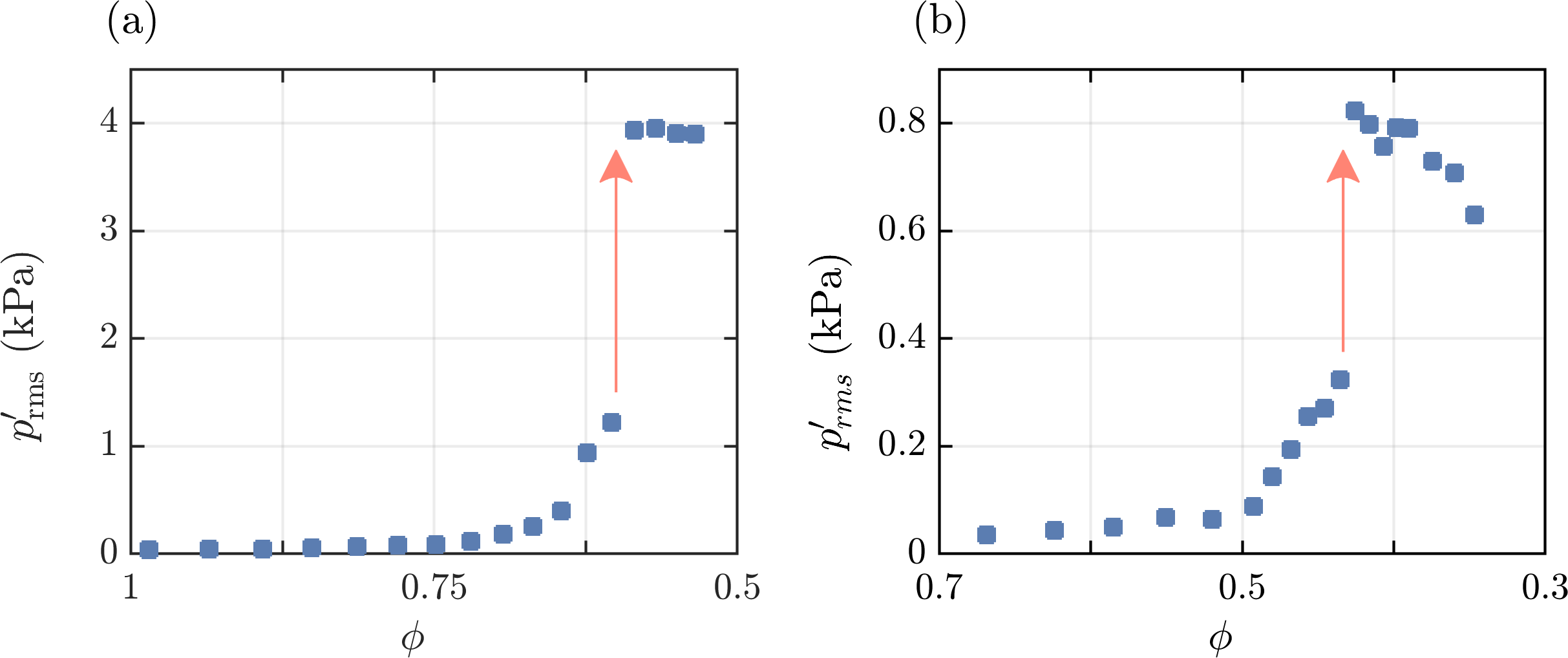}
\caption{The variation of $p'_\mathrm{rms}$ as a function of $\phi$ during secondary bifurcation in (a) the swirl-stabilized dump combustor and (b) the bluff-body stabilized dump combustor with preheated air.}
\label{Experimentalresults}
\end{figure}

Figure \ref{Experimentalresults} shows the characteristics of abrupt transition to large amplitude limit cycle oscillation in the swirl-stabilized and preheated bluff-body stabilized combustor as $\phi$ is decreased. The transition is observed when $\phi$ is decreased from 0.99 to 0.54 in a quasi-static manner in the swirl-stabilized combustor, while it is observed for a decrease in $\phi$ from 0.67 to 0.35 in the preheated bluff-body stabilized combustor. In each of these two combustors, a decrease in $\phi$ leads to a transition from combustion noise to high-amplitude TAI through the states of intermittency and low-amplitude limit cycle oscillations. These states in the swirl and bluff-body combustor have similar statistical properties to the representative plots shown in Fig. \ref{Annualr timeseries}(b-e). The abrupt transition, thus, takes place through a secondary bifurcation to large amplitude limit cycle oscillations. Note that the swirl-stabilized dump combustor depicts secondary bifurcation to very large amplitude levels ($p'_\text{rms}\approx 4$ kPa). 

Thus, it is evident from Figs. \ref{Annualr timeseries} and \ref{Experimentalresults} that these turbulent thermoacoustic systems exhibit abrupt transitions in the form of secondary bifurcations. Secondary bifurcation appears in disparate turbulent combustion systems with very different flame and acoustic responses. Thus, the common phenomenology across disparate combustors implies a certain universal mechanism through which secondary bifurcation occurs in turbulent combustors. Motivated by these results, we consider the model for describing secondary bifurcations proposed by \citet{ananthkrishnan1998application}. We extend the model to obtain primary and secondary limit cycle solutions, derive the underlying potential functions, and underscore the role of stochastic fluctuations on the observed phenomenology. 

\section{Modeling secondary bifurcation in thermoacoustic systems} 
\label{Thermoacoustic model of secondary bifurcation}
\subsection{Low-order thermoacoustic model} \label{model to describe secondary bifurcations}
In the above experiments, we observed secondary bifurcation associated with the longitudinal mode of TAI in three different combustors. Since we are concerned with modeling this transition, we consider a simplified one-dimensional thermoacoustic system where the longitudinal mode is excited. We neglect the effects of temperature gradient and mean flow. Further, pressure fluctuations relative to the mean are not very large, so the nonlinear acoustic terms are unimportant. We will see that these are reasonable assumptions for determining the characteristic features of abrupt secondary bifurcations. 

Thus, the flame-acoustic interactions inside the combustor are governed by the linearized momentum and energy conservation equations \cite{nicoud2009zero}, which are given as,
\begin{align}
    \frac{1}{\bar\rho} \frac{\partial p^\prime(z,t) }{\partial  z} + \frac{\partial u^\prime(z,t) }{\partial  t} &= 0,\label{ET1}\\
    \frac{\partial p^\prime(z,t) }{\partial  t} + \gamma \bar p \frac{\partial u^\prime(z,t) }{\partial  z} &= (\gamma -1) \dot{ Q}^\prime(z,t) \delta( z-{z_f}).\label{ET2}
\end{align}
Here, $\gamma$ is the ratio of specific heat capacities, $t$ is time, and $z$ is the distance along the axial direction of the duct. $u^\prime$ and $p^\prime$ are the velocity and pressure fluctuations, while $\bar{p}$ and $\bar{\rho}$ indicate the mean pressure and density, respectively. We assume the flame to be compact such that the heat release rate fluctuations  $\dot{Q}'$ are concentrated at the flame location $z_f$, indicated by the Dirac-delta ($\delta$) function \cite{mcmanus1993review}. Equations (\ref{ET1}) and (\ref{ET2}) can be suitably modified to yield an inhomogeneous wave equation with the source term due to the heat release rate fluctuations, as given below \cite{lieuwen2021unsteady}: 
\begin{equation}\label{ET5}
     c^2 \frac{\partial^2 p'(z,t)}{\partial z^2} - \frac{\partial^2 p'(z,t)}{\partial t^2} = -(\gamma -1) \frac{\partial \dot{ Q}'(z,t)}{\partial t} \delta( z-{z_f}),
\end{equation}
where, $c=\sqrt{\gamma\bar{p}/\bar{\rho}}$ is the speed of sound.

We use a Galerkin modal expansion to simplify the second-order partial differential equation into an ordinary differential equation in the time domain \cite{lores1973nonlinear}. The acoustic velocity and pressure fluctuations are projected on a set of spatial basis functions (cosines and sines) having temporal coefficients ($\eta, \dot{\eta}$), respectively, as indicated below:
\begin{equation}\label{ET6}
    p'( z,t)= \bar p \sum_{j=1}^{n} \frac{\dot\eta_j(t)}{\omega_j} \cos(k_j z)  \quad\text{and}\quad   u'( z,t)= \frac{\bar p}{\bar\rho  c} \sum_{j=1}^{n} \eta_j(t) \sin(k_j z),
\end{equation}
where $j$ represent the eigenmodes. The basis functions chosen here are orthogonal, satisfy the acoustic (closed-open) boundary conditions, and form the eigenmodes of the self-adjoint part of the linearized equations \cite{balasubramanian2008thermoacoustic}. Here, $k_j$ is the wavenumber given by $k_j = (2j-1) \pi /2 L $, where $L$ is the length of the combustor. The relationship between the wavenumber and the natural frequency of each mode can be expressed as $\omega_j =  c k_j$. Substituting Eq. (\ref{ET6}) into Eq. (\ref{ET5}) yields, 
\begin{equation}\label{ET7}
    \sum_{j=1}^{n} \frac{\ddot \eta_j(t)}{\omega_j} \cos{(k_j z)}+\frac{\gamma \bar p}{\bar \rho  c} \sum_{j=1}^{n} \eta_j(t) k_j \cos(k_j z) = \frac{\gamma - 1}{\bar p} \dot Q' \delta(z-z_f).
\end{equation}
Evaluating the inner product of Eq. (\ref{ET7}) along each of the basis function and integrating the resulting equation over the length of the combustor we get,
\begin{equation}\label{ET8}
    \frac{\ddot\eta_j(t)}{\omega_j}+  c k_j \eta_j(t) = \frac{2(\gamma -1)}{L \bar p} \int_{0}^{L} \dot Q' \delta(z-z_f) \cos(k_j z) dz.
\end{equation}

Here, we restrict our analysis to a single eigenmode which is sufficient for analyzing transition characteristics observed in the combustors described here. Further, the observed dynamics in thermoacoustic systems arise from the nonlinear flame-response to acoustic perturbations. So, the heat release rate response can be expressed as a nonlinear function of $\eta$ and $\dot\eta$, i.e, $Q'\equiv Q'(\eta, \dot\eta)$. With all these considerations, Eq. (\ref{ET8}) reduces to the equation of a self-excited harmonic oscillator, expressed as
\begin{equation}\label{ET16}
    \ddot{\eta} + \omega^2 \eta = f(\eta, \dot{\eta}),
\end{equation}
where, $f(\eta,\dot \eta)=\dot Q^\prime(\eta,\dot{\eta}) - \alpha \dot \eta$ is the nonlinear driving term with an extra term $\alpha \dot \eta$ added to take acoustic damping into account ($\alpha$ is the damping co-efficient). Using a truncated Taylor series expansion of the source term $f(\eta,\dot \eta)$ following \cite{culick2006unsteady,lieuwen2003statistical}, we express Eq. (\ref{ET16}) as  
\begin{equation}\label{sbeqn}
\ddot{\eta}+\left ( \mu_6 \eta^6+ \mu_4 \eta^4+\mu_2 \eta^2- \mu_0 \right ) \dot{\eta}+\omega^2 \eta+\xi=0,
\end{equation}
where $\mu_0$ is the control parameter and $\mu_2$, $\mu_4$, and $\mu_6$ are the coefficients of the nonlinear terms. The effect of turbulence is included as additive Gaussian white noise $\xi$ which is delta correlated in time: $\langle \xi(t) \xi(t+\tau)\rangle = \Gamma \delta (\tau)$, where $\Gamma$ is the noise intensity \cite{noiray2013deterministic, noiray2017method, bonciolini2017subcritical}. The symbol $\left \langle . \right \rangle$ represents the ensemble of realizations of the stochastic process. Thus, the overall dynamics is governed by the second-order stochastic differential equation. 

\begin{figure}
\centering
\includegraphics[width=\textwidth]{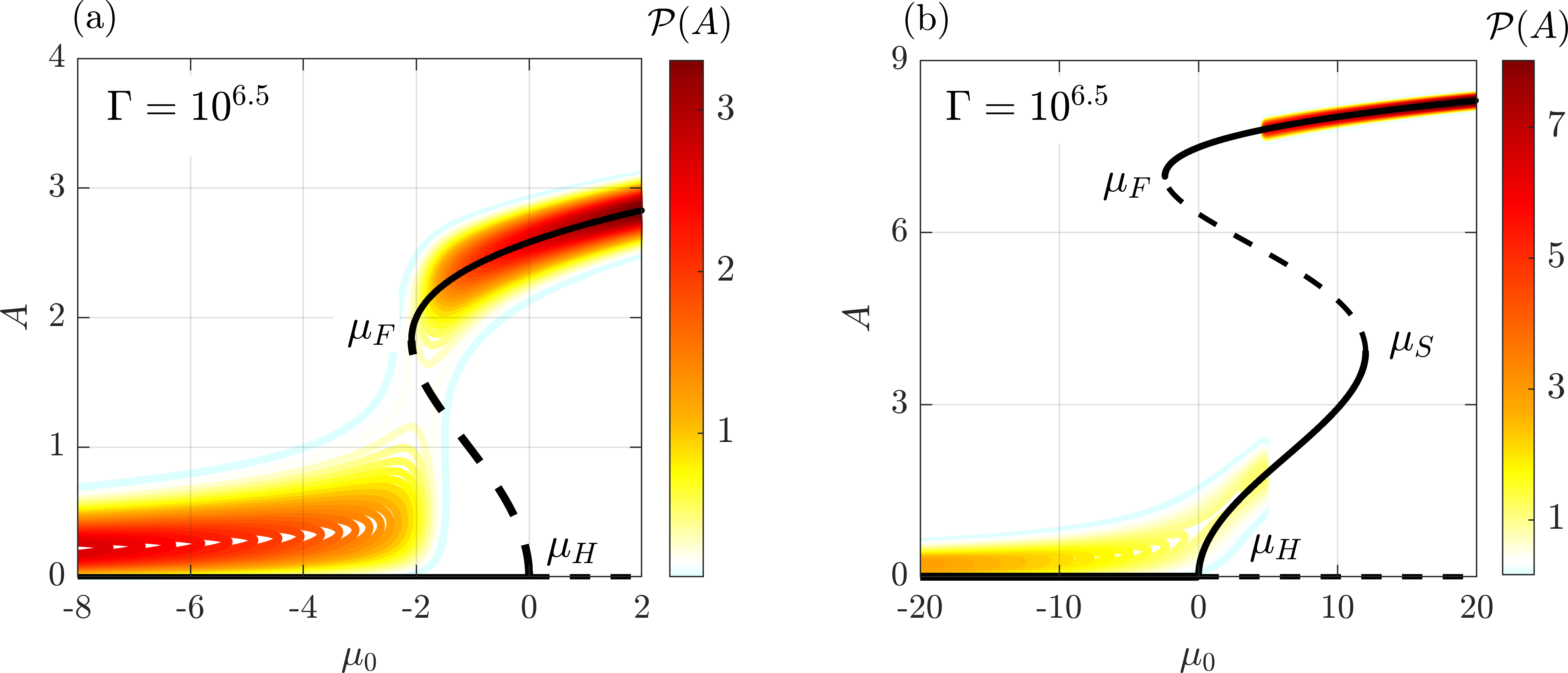}
\caption{Bifurcation characteristics of the stochastic thermoacoustic system described by Eq. (\ref{sbeqn}). Variation in the amplitude of fluctuations for (a) subcritical and (b) supercritical bifurcation followed by a secondary bifurcation to large amplitude limit cycle. The bifurcation diagram for the deterministic system ($\Gamma=0$, cf. Eq. \ref{Eq:AmpLangevin}) is indicated by the black line. The difference in the abscissa in (a) and (b) indicates the significant difference between the amplitude of limit cycles due to subcritical and secondary bifurcation. The solid lines correspond to stable solution, while the broken lines indicate the unstable solution. The contour shows the variation in probability density function $\mathcal{P}(A)$ with parameter $\mu_0$, estimated according to Eq. \eqref{E_pdf}. The noise intensity is fixed at $\Gamma=10^{6.5}$. The other model parameters are: (a) $\mu_2 = -10$, $\mu_4 = 3$, $\mu_6 = 0$; and (b) $\mu_2 = 7$, $\mu_4 = -0.6$, $\mu_6 = 0.01$. Labels $\mu_H$, $\mu_F$ and $\mu_S$ indicate the parameter value $\mu_0$ at which Hopf, fold and secondary bifurcations occur.}
\label{Fig4-SubSec_Bif}
\end{figure}

\subsection{Secondary bifurcation of the deterministic system} \label{secondary bifurcation of the deterministic system}
Let us first consider the deterministic system ($\Gamma=0$). It is instructive to examine the effect that each of the nonlinear terms has on the overall dynamics of the system. The higher-order nonlinear terms ($\eta^2,~\eta^4$ and $\eta^6$) in the expansion of $f(\eta,\dot \eta)$ are responsible for stabilizing the system to limit cycle solution when the fixed point solution becomes unstable due to a variation in $\mu_0$ \cite{ananthkrishnan1998application}. To see this, set $\mu_6=\mu_4=0$ and $\mu_2=1$. The system depicts a stable fixed point for $\mu_0<0$. The fixed point solution becomes unstable at $\mu_0=0$, and the eigenvalues representing the linearized system cross the imaginary axis. This is the well-known scenario of supercritical Hopf bifurcation to limit cycle oscillations (not shown here for brevity). Near the fixed point $\mu_H$, the amplitude of the limit cycle increases monotonically through the relation $\eta\sim(\mu_0-\mu_H)^{0.5}$ \cite{strogatz2018nonlinear}. 

The stable limit cycle so obtained can be made unstable by choosing a negative value for $\mu_2$. This is depicted by the broken line in Fig. \ref{Fig4-SubSec_Bif}(a) (shown for the representative case of $\mu_2=-10$), which has a parabolic dependence close to $\mu_H$. The unstable limit cycle solution can be stabilized by the inclusion of the next higher-order term $\eta^4$ by setting $\mu_4=3$. The inclusion of $\eta^4$ stabilizes the limit cycle through a fold bifurcation ($\mu_0=\mu_F$), and the system depicts large amplitude limit cycle oscillation. We observe that the system exhibits two stable solutions for $\mu_F<\mu_0<\mu_H$: a fixed point and a limit cycle. Thus, the system is said to exhibit bistability. With $\mu_6=0$, Eq. \eqref{sbeqn} reverts to the normal form of subcritical Hopf bifurcation and is widely used as models of subcritical bifurcation in the thermoacoustic literature \cite{moeck2008subcritical, laera2017finite, bonciolini2017subcritical}.

We obtain secondary bifurcation through the inclusion of the sixth-order nonlinear term $\eta^6$. Figure \ref{Fig4-SubSec_Bif}(b) shows the typical bifurcation plot obtained upon setting $\mu_2=7, \mu_4=-0.6$ and $\mu_6=0.01$. The system depicts a stable fixed point state for $\mu_0<0$. At $\mu_0=\mu_H=0$, the fixed point becomes unstable, and a supercritical bifurcation to limit cycle solution takes place. There is a monotonic increase in $A$ with $\mu_0$ until the system reaches the secondary bifurcation point $\mu_0=\mu_S=12$. At $\mu_S$, the supercritical branch becomes unstable, and there is an abrupt jump to a stable large amplitude limit cycle solution. 

In the reverse direction, the secondary limit cycle branch extends till the fold bifurcation with $\mu_F<\mu_H$, highlighting the associated bistability in the system. Further, there are two regions of bistability. First, in the region $\mu_F<\mu_0<\mu_H$, the system depicts a fixed point solution and secondary limit cycle. Second, for $\mu_H<\mu_0<\mu_S$, the system depicts a low-amplitude limit cycle solution and secondary limit cycle solution. This behavior (also referred to as hysteresis) is qualitatively very similar to the bistable region observed in our experiments (cf. Fig. \ref{Annualr timeseries}a). 

\subsection{Slow flow representation of the stochastic system}\label{Slow flow representation of the stochastic system}
\label{Role of turbulent fluctuations}
Let us now consider the effect of stochastic fluctuations on the transition to limit cycle oscillations. We consider the acoustic variable $\eta(t)$ to be quasi-harmonic \cite{minorsky1962nonlinear}, such that we have: 
\begin{equation}\label{EKB1}
    \eta(t)= A(t) \cos\left[\omega t+\phi(t)\right].
\end{equation}
This decomposition allows us to separate the evolution of envelope-amplitude $A(t)$ and phase $\phi(t)$, which vary at a slower time scale in comparison to the faster time scale $2\pi/\omega$. Now, we evaluate the expressions for $\eta, \dot\eta$ and $\ddot\eta$ after expressing Eq. (\ref{EKB1}) in terms of the exponential function (refer to \ref{appendix: section1}). Substituting these in Eq. (\ref{sbeqn}) leads to
\begin{equation}\label{EKB2}
 i \omega \dot{a} e^{i \omega t}  - \frac{\omega^2}{2} \beta + \left (\frac{\mu_6}{64} \beta^6 + \frac{\mu_4}{16}\beta^4 + \frac{\mu_2}{4}\beta^2- \mu_0 \right)
    \times  \frac{i \omega}{2} (a e^{i \omega t} - a^* e^{-i \omega t}) + \frac{\omega^2}{2} \beta + \xi= 0,
\end{equation}
where, $\beta=a e^{i \omega t}+ a^* e^{-i \omega t}$, $a=A e^{i \phi}$, $ a^*=A e^{-i \phi}$. To eliminate the fast time scale, we compute the average of Eq. (\ref{EKB2}) over the time period $T=2\pi/\omega$ of the fast oscillations \cite{balanov2009simple}. Using the method of averaging and simplifying the stochastic functions following \cite{krylov2016introduction, stratonovich1967topics,balanov2009simple} (refer to \ref{appendix: section2} for further details), we obtain a set of Langevin equation governing the evolution of the slowly varying amplitude and phase of the system, which are expressed as:
\begin{align}
    \dot{A} &=  \frac{\mu_0}{2}A-\frac{\mu_2}{8}A^3-\frac{\mu_4}{16}A^5-\frac{5 \mu_6}{128}A^7+\frac{\Gamma}{4\omega^2 A}+\frac{\sqrt{\Gamma}}{\sqrt{2}\omega}\xi_1, \label{Eq:AmpLangevin}\\
    \dot{\phi} &=\frac{\sqrt{\Gamma}}{\sqrt{2}\omega A}\xi_2.\label{Eq:PhaseLangevin}
\end{align}
Here, $\xi_1$ and $\xi_2$ are two uncorrelated Gaussian white noise terms with zero mean and unit variance. Note that for a deterministic system, the evolution of the phase is zero ($\dot \phi = 0$), and now with the addition of noise, the phase drifts.
In Eq. \eqref{Eq:AmpLangevin}, the sign associated with the factor $\mu_0/2$ determines the linear stability of the system. Further, the term $\Gamma/4\omega^2 A+(\sqrt{\Gamma}/\sqrt{2}\omega)\xi_1$ arises due to the covariance of stochastic terms in the Fokker-Planck equation (refer to Eq. \ref{E7.46}) of the joint PDF of $A$ and $\phi$. The Langevin equation (Eq. \ref{Eq:AmpLangevin}) can be expressed in terms of the potential function $V$, as shown below: 
\begin{equation} \label{E_potentialdefination}
    \dot{A}=-\frac{dV}{dA}+\frac{\sqrt{\Gamma}}{\sqrt{2}\omega}\xi_1.
\end{equation} 
Here, the negative sign associated with it implies the fact that the evolution of the system tends to minimize the potential function. The potential function $V(A,\mu_0)$ can then be determined by comparing Eqs. \eqref{Eq:AmpLangevin} and \eqref{E_potentialdefination} and evaluating the resulting integral. This leads to 
\begin{equation}\label{Epotential_s}
    V(A) = - \frac{\mu_0}{4}A^2+\frac{\mu_2}{32}A^4+\frac{\mu_4}{96}A^6+\frac{5\mu_6}{1024}A^8-\frac{\Gamma}{4\omega^2}\ln{A},
\end{equation}
which defines the potential function of the overall system.

\subsection{Stationary solution of the Fokker-Planck equation}\label{Stationary solution of the Fokker-Planck equation}
We now recast the stochastic differential equation in the \^Ito sense \cite{Grigorios2014stochastic}, which would allow us to invoke the Fokker-Planck equation for the evolution of $\mathcal{P}(A)$ corresponding to the Langevin equation of $A$. Thus, in the \^Ito sense, Eq. (\ref{E_potentialdefination}) becomes
\begin{equation}
    dA = \Psi(A)dt+dW,
\end{equation}
where,
\begin{equation}\label{Eamp_with_noise}
\Psi(A) = -\frac{dV}{dA}=\frac{\mu_0}{2}A -\frac{\mu_2}{8}A^3 -\frac{\mu_4}{16}A^5 -\frac{5 \mu_6}{128}A^7 +\frac{\Gamma}{4\omega^2A},
\end{equation}
and $dW=\xi dt$ is the increment of the Wiener process. Recall that we have assumed that the noise $\xi$ is delta-correlated, a condition that is seldom fulfilled in real systems. The noise usually possess finite correlation time ($t_{\text{cor}}$). For the present purposes, if $\xi$ is sufficiently fast such that $t_{\text{cor}}$ is much lesser than the relaxation time of the system, the evolution of the PDF $\mathcal{P}(A)$ satisfies the Fokker-Planck equation
\begin{equation}
\frac{\partial}{\partial t}\mathcal{P}(A,t)=-\frac{\partial}{\partial A}[\Psi(A) \mathcal{P}(A,t)]+\frac{\Gamma}{4\omega^2} \frac{\partial^2}{\partial A^2} \mathcal{P}(A,t).
\label{Eq-FPEqn}
\end{equation}
Here, $\Psi(A)$ and $\Gamma/4\omega^2$ are the drift and diffusion coefficients, respectively. At large times, we assume that the distribution reaches a stationary state, such that: $\lim_{t\rightarrow \infty} \mathcal{P}(A,t) =\mathcal{P}(A)$. Thus, Eq. \eqref{Eq-FPEqn} reduces to
\begin{equation}
    \frac{d}{dA}\mathcal{P}(A)-\frac{4\omega^2}{\Gamma}\Psi(A)\mathcal{P}(A)=0.
\end{equation}
This equation can be readily solved to yield
\begin{equation}\label{E_pdf}
\mathcal{P}(A) = \mathcal{N} \exp\left( -\frac{4\omega^2}{\Gamma}V(A) \right).
\end{equation}
where $\mathcal{N}$ is a constant such that $\int_0^{\infty} \mathcal{P}(A)=1$. 

\subsection{Effect of noise levels on abrupt transitions}\label{Effect of noise levels on the transition}
In Fig. \ref{Fig4-SubSec_Bif}, we plot the analytically derived probability distribution function $\mathcal{P}(A)$ in Eq. (\ref{E_pdf}) as a function of $\mu_0$. For obtaining the subcritical bifurcation, we set $\mu_6 = 0, \mu_4 = 3$ and $\mu_2 = -10$, while for the secondary bifurcation, we choose $\mu_6 = 0.01, \mu_4 = -0.6$ and $\mu_2 = 7$. The bifurcation for the purely deterministic case is obtained by setting the noise intensity $\Gamma=0$ in Eq. (\ref{Eq:AmpLangevin}) and plotting the resulting solution. Figure \ref{Fig4-SubSec_Bif} compares the effect of the same level of noise intensity $\Gamma=10^{6.5}$ on the characteristic of bifurcation. We notice that for the same noise intensity, an initially sub-critical Hopf bifurcation transforms into a continuous sigmoid-type transition, as depicted by the contour of $\mathcal{P}(A)$ in Fig. \ref{Fig4-SubSec_Bif}(a). On the other hand, the secondary bifurcation remains abrupt with an important difference: the fixed point solution is colored by noisy fluctuations, which hides the sharp demarcation in the dynamics at the location of supercritical transition $\mu_0=\mu_H$. This is precisely what we observe in the bifurcation plots from experiments (cf. Figs. \ref{Annualr timeseries}a and \ref{Experimentalresults}) where the amplitude rises through the state of intermittency before the state of the low-amplitude limit cycle is reached.

\begin{figure}
\centering
\includegraphics[width=\textwidth]{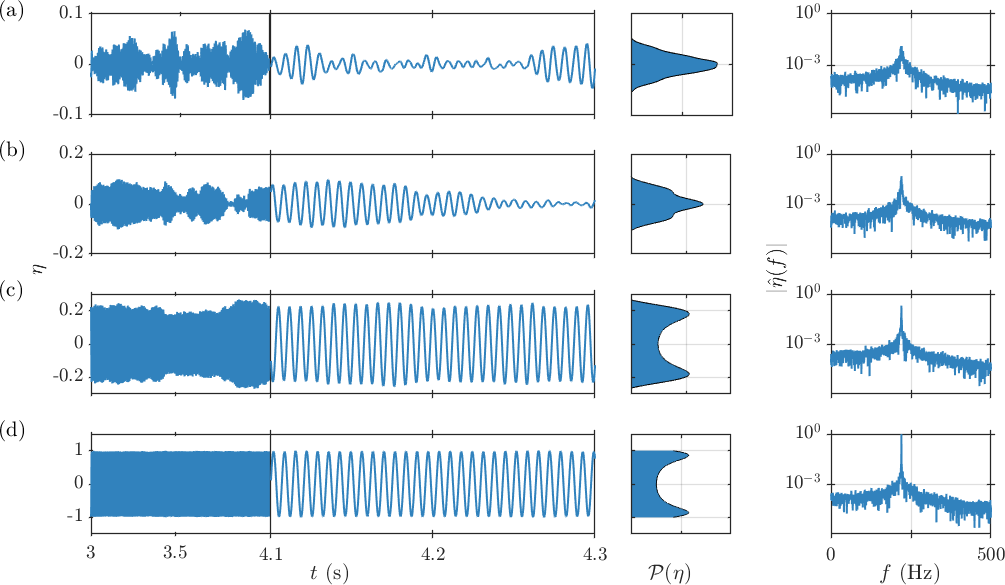}
\caption{Time series of $\eta(t)$, the probability density function $\mathcal{P}(\eta)$ and the spectrum $\hat{\eta}(f)$ from the stochastic model during the state of (a) combustion noise ($\mu_0=-20$), (b) intermittency ($\mu_0=-5$), (c) low-amplitude TAI ($\mu_0=5$) and (d) high-amplitude TAI ($\mu_0=12$). The simulation parameters are: $\mu_6 = 0.01$, $\mu_4 = -0.6$, $\mu_2 = 7$ and $\Gamma=10^{6.5}$. Note that $\eta$ is normalized by the amplitude of limit cycle oscillation shown in panel (d).}
\label{Fig:ModelTimeSeries}
\end{figure}

To see this effect clearly, we numerically simulate the model (Eq. \ref{sbeqn}) using the stochastic Runge-Kutta method and plot the time series, $\mathcal{P}(\eta)$ and $\hat{\eta}(f)$ for four representative states across the transition in Fig. \ref{Fig:ModelTimeSeries}. At $\mu=-20$, the time series depicts aperiodic fluctuations, albeit with some periodic content (cf. Fig. \ref{Fig:ModelTimeSeries}a). However, the spectral amplitude remains very low. The distribution $\mathcal{P}(\eta)$ shows unimodal characteristics, a fact also observed in the experimental data (cf. Fig. \ref{Annualr timeseries}a). We also note here that features such as chaos \cite{tony2015detecting,nair2013loss} and multifractality \cite{nair2014intermittency} of the state of combustion noise are not captured by the additive white noise considered here. Next, at $\mu_0=-5$ (cf. Fig. \ref{Fig:ModelTimeSeries}b), we observe intermittent bursts amidst aperiodic fluctuations. The distribution $\mathcal{P}(\eta)$ shows a change from a unimodal distribution to peaks at $|\eta|\neq 0$, a feature we also noted in Fig. \ref{Annualr timeseries}(c). Finally, we observe low-amplitude and high-amplitude limit cycle oscillations at $\mu_0=5$ and $\mu_0=12$ as shown in Fig. \ref{Fig:ModelTimeSeries}(c,d). 


We next quantify the effect of turbulence on the transition to the final state of limit cycle oscillations. To do this, we define a transition amplitude factor $R$ from experiments ($R_E$) and model ($R_M$) as
\begin{equation}
    R_E=\frac{p'_{\text{rms(CN)}}}{p'_{\text{rms(LCO)}}-p'_{\text{rms(CN)}}}\times 100 \%, \qquad R_M=\frac{\eta_{\text{rms(CN)}}}{\eta_{\text{rms(LCO)}}-\eta_{\text{rms(CN)}}}\times 100 \%.
\end{equation}
The transition amplitude factor compares the amplitude of combustion noise and the difference in the amplitude of combustion noise and limit cycle from experiments and that simulated in the model. Thus, the transition amplitude factor $R_E$ can be used to quantify the effect of turbulence on the observed transition and compared with representative values of $R_M$ for a given intensity $\Gamma$, to understand how well the effect of turbulent fluctuations are approximated by the stochastic model (see Table \ref{Tab1}). 

The transition amplitude factor during the abrupt transition in the annular, swirl-stabilized, and bluff-body stabilized dump combustor are $R_E=1.4\%$, $2.3\%$, and $8.4\%$, respectively. For each of these cases, the maximum Reynolds numbers attained during the transitions are $8.6\times10^3$, $2.6\times10^4$, and $5.67\times 10^4$, respectively. The values of $R_E$ obtained from the experiments are comparable to the values of $R_M=9\%$ obtained from the model, for a representative value of intensity $\Gamma = 10^{6.5}$, during the secondary bifurcation, which is illustrated in Fig. \ref{Fig4-SubSec_Bif}(b). In contrast, when we consider the case of initially subcritical bifurcation made continuous by the additive noise, as shown in Fig. \ref{Fig4-SubSec_Bif}(a), the factor is $R_M=22\%$. To put this into context, the factor for the intermittent transition reported in Fig. 4(c) by \citet{nair2014intermittency}, is $R_E=21\%$. Thus, the effect of turbulence on the observed transition is approximated reasonably well in the stochastic model. 

To further the discussion, we now set the noise intensity in the model to $\Gamma = 10^{7.5}$. Fig. \ref{Fig5-NoiseSec_Bif}(b) shows the stochastic bifurcation for this case. We note that the secondary bifurcation displays a continuous, sigmoid-type transition. This noise level corresponds to the transition amplitude factor of $R_M=35\%$. Thus, we infer that the abrupt secondary bifurcations can be made continuous at very high noise levels, which may be unrealizable in practical turbulent combustors. 

\begin{table}[]
\resizebox{\textwidth}{!}{%
\begin{tabular}{llllllll}
\hline
\multicolumn{4}{c|}{Abrupt transitions}& \multicolumn{4}{c}{Continuous transitions}\\ \hline
\multicolumn{2}{c|}{Experiments} & \multicolumn{2}{c|}{Model} & \multicolumn{2}{c|}{Experiments} & \multicolumn{2}{c}{Model}\\ \hline
\begin{tabular}[c]{@{}l@{}} Annular\\ combustor\\ (Fig. \ref{Annualr timeseries}a) \\ ~\end{tabular}                 & \multicolumn{1}{c|}{$R_E=1.4\%$} & \multirow{3}{*}{\begin{tabular}[c]{@{}l@{}}Stochastic \\ secondary\\ bifurcation\\ (Fig. \ref{Fig4-SubSec_Bif}b, \\ at $\Gamma = 10^{6.5}$)\end{tabular}} & \multicolumn{1}{l|}{\multirow{3}{*}{$R_M=9\%$}} & \multirow{3}{*}{\begin{tabular}[c]{@{}l@{}}Continuous\\ transition in \\ \citet{nair2014intermittency} \end{tabular}} & \multicolumn{1}{l|}{\multirow{3}{*}{$R_E=21\%$}} & \multirow{3}{*}{\begin{tabular}[c]{@{}l@{}}Stochastic\\ subcritical\\ bifurcation\\ (Fig. \ref{Fig4-SubSec_Bif}a, \\ at $\Gamma = 10^{6.5}$ )\end{tabular}} & \multirow{3}{*}{$R_M=22\%$} \\
\begin{tabular}[c]{@{}l@{}}Swirl \\ dump combustor\\ (Fig. \ref{Experimentalresults}a) \\ ~\end{tabular}                 & \multicolumn{1}{l|}{$R_E=2.3\%$} & & \multicolumn{1}{l|}{} & & \multicolumn{1}{l|}{} & & \\
\begin{tabular}[c]{@{}l@{}}Preheated\\ bluff-body \\ dump combustor\\ (Fig.\ref{Experimentalresults}b)\end{tabular} & \multicolumn{1}{l|}{$R_E=8.4\%$} & & \multicolumn{1}{l|}{} & & \multicolumn{1}{l|}{}  & & \\ \hline
\end{tabular}
}
\caption{Comparison of the transition amplitude factor observed in experiments $R_E$ and simulated from the stochastic model $R_M$ for representative values of noise intensity $\Gamma$.}
\label{Tab1}
\end{table}

\subsection{Potential landscape of the secondary bifurcation}
\label{Potential landscape of the secondary bifurcation}
Let us now consider the stability of the dynamical states observed during the secondary bifurcation. The stability of various dynamical states is best visualized through the potential function $V$ (Eq. \ref{Epotential_s}). From the definition of the potential function, we have: $\Psi(A)=-\partial V/\partial A$. We note that minima and maxima of the potential function $V(A,\mu_0)$ correspond to the stable and unstable fixed points of $\Psi(A, \mu_0)$. The value of ${\Psi^\prime(A)}$ is a measure of the stability of the fixed points as $\Psi^\prime(A)$ corresponds to the second derivative of potential function ($d^2V/dA^2$) \cite{strogatz2018nonlinear}. The second derivative is a measure of the curvature of the potential function, describing its sharpness. Thus, the higher the magnitude of $-{\Psi^\prime(A)}$ is, the higher would be the stability of the fixed point \cite{strogatz2018nonlinear}. We discuss how the variation in the parameter $\mu_0$ leads to a change in the stability of the potential $\Psi(A)$ next.

\begin{figure}
\centering
\includegraphics[width=\textwidth]{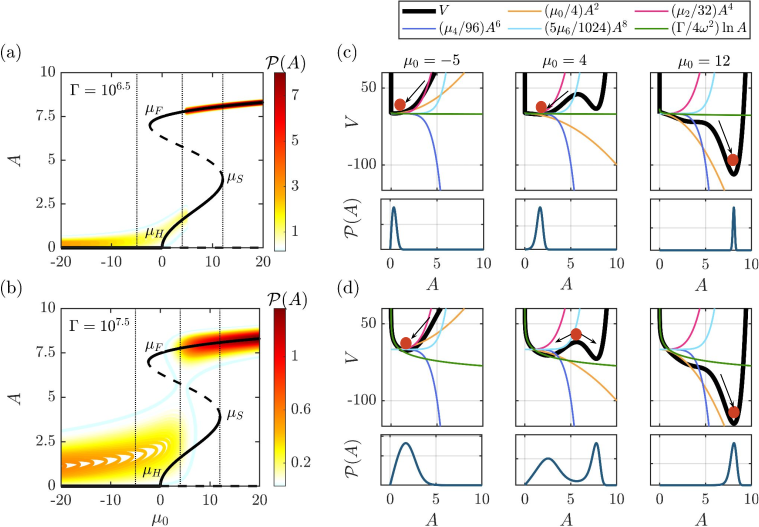}
\caption{Effect of stochastic fluctuations on the properties of secondary bifurcation for (a) $\Gamma=10^{6.5}$ and (b) $\Gamma=10^{7.5}$. The contours show the variation in the distribution $\mathcal{P}(A)$ as a function of $\mu_0$. Panels (c) and (d) depicts the potential functions $V(A)$ (top panel) and distribution $\mathcal{P}(A)$ (bottom panel). The potential $V$ is indicated by the bold line. The contributions of individual terms of Eq.~\eqref{Epotential_s} are also indicated. $V(A)$ and $\mathcal{P}(A)$ are shown at $\mu_0 = -5,~4$ and 12, marked by the dotted lines in panel (a) and (b). Other simulation parameters are same as that in Fig.~\ref{Fig4-SubSec_Bif}(b).}
\label{Fig5-NoiseSec_Bif}
\end{figure}

The variation in the potential function $V(A)$ is shown in Fig. \ref{Fig5-NoiseSec_Bif}(c,d) at $\mu_0=-5,~4$ and 12 to compare their behavior at different states, indicated by the dotted lines in the transition diagram. The potential functions are shown at two different noise intensities ($\Gamma=10^{6.5}$ and $\Gamma=10^{7.5}$). The associated distribution $\mathcal{P}(A)$ is also shown below the potential $V(A)$. When $\mu_0=-5$, the system is at stable equilibrium (Fig. \ref{Fig5-NoiseSec_Bif}c), and any amount of perturbation to the stable state will be restored to its equilibrium position. Thus, the system exhibits globally stable fixed points for $\mu_0<0$. The additive noise continuously perturbs the system around the stable fixed points. For low noise levels ($\Gamma = 10^{6.5}$), the mean of the distribution $\mathcal{P}(A)$ is at the minimum of the potential function $V(A)$ (cf. Fig. \ref{Fig5-NoiseSec_Bif}c, at $\mu_0 = -5$). In contrast, at a higher noise level ($\Gamma = 10^{7.5}$), $\mathcal{P}(A)$ shows a wider distribution, as the variance of the noise is much larger (see Fig. \ref{Fig5-NoiseSec_Bif}(d), at $\mu_0 = -5$). 
\begin{figure}
\centering
\includegraphics[width=0.5\textwidth]{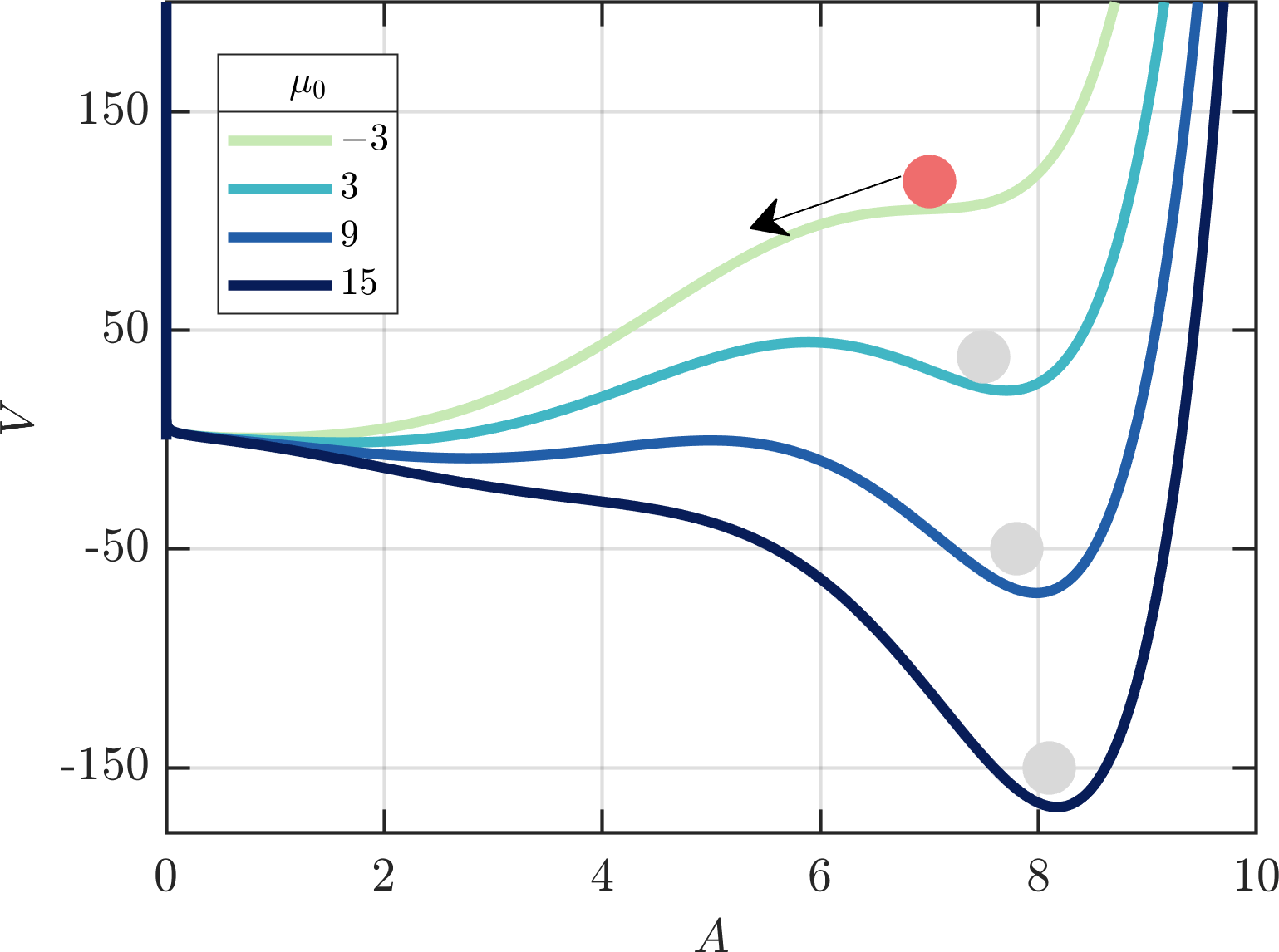}
\caption{Illustration of hysteresis observed during the secondary bifurcation. Plot shows the variation of the potential function $V(A)$ when the value of the parameter is decreased from $\mu_0 = 15$ to $\mu_0 = -3$.}
\label{hysteresisfigure}
\end{figure}
 
For $\mu_0>0$, we notice that the potential $V(A)$ develops a secondary trough. We observe a unimodal Gaussian distribution $\mathcal{P}(A)$ for $\Gamma=10^{6.5}$ at $\mu_0 = 4$ and $\mu_0=12$ (Fig. \ref{Fig5-NoiseSec_Bif}c). We notice that there is a shift in $\mathcal{P}(A)$ from a distribution centered at a lower amplitude to one centered at a much higher amplitude. This shift at  $\mu_0=12$ is associated with the secondary trough in the potential $V(A)$ becoming the global minima, implying its global stability. On the other hand, for $\Gamma = 10^{7.5}$, we notice that at $\mu_0=4$, the potential $V(A)$ has a double-well characteristic with a comparable value of minima. As the noise level is higher, the perturbations can take the system from one potential well to the next. Consequently, the distribution $\mathcal{P}(A)$ has a bimodal distribution. The bimodal distribution implies the presence of intermittency where the amplitude switches between low-amplitude oscillations (wider distribution) and higher-amplitude bursts of periodic oscillations (narrow distribution). Finally, at $\mu_0=12$, we have a unimodal distribution. The second trough has a lower minima, implying that the system reaches the globally stable limit cycle oscillation at very large amplitude levels.

To illustrate the phenomenon of hysteresis, we plot the potential functions for different values of $\mu_0 $ (15, 9, 3, and -3), which is shown in Fig. \ref{hysteresisfigure}. The system depicts the state of the secondary limit cycle at $\mu_0 = 15$. Reversing $\mu_0$ below $\mu_S$ will not restore the state of the system back to the primary limit cycle. Upon reducing the parameter value ($\mu_0=9$ and 3), another trough forms at a lower amplitude value. However, the potential barrier (local maxima in $V$) of the unstable fixed point separating the two troughs is very high, thereby hindering the transitions back to either the primary limit cycle or the fixed point. The system transition to the fixed point solution only when the parameter is reversed below the fold point (i.e., $\mu_0<\mu_F$). 

\section{Conclusion} \label{Conclusion}
To summarize, we reported the observation of secondary bifurcation in three disparate turbulent combustors -- annular combustor, swirl-stabilized combustor, and bluff-body stabilized combustor with preheated air -- despite them having completely different flame responses and acoustic characteristics. These systems exhibit a sequence of transitions from combustion noise to intermittency to low-amplitude limit cycle oscillations, followed by an abrupt jump to large amplitude secondary limit cycle oscillations. We then model the secondary bifurcation using a second-order oscillator equation containing higher-order nonlinearities. The effect of turbulence is incorporated in terms of Gaussian delta-correlated white noise. We show that the model captures the secondary bifurcation very well and depicts good qualitative agreement with the dynamical states observed in experiments.

We then derive the Langevin equation of the slowly varying amplitude and phase through deterministic and stochastic averaging techniques. We obtain the potential function for the secondary bifurcation and discuss the stability of the observed dynamical states. In addition, we obtain the stationary distribution of the envelope of the amplitude of the fluctuations by solving the Fokker-Plank equation. 
We show that a deterministic subcritical bifurcation is transformed into a continuous sigmoid type transition, typical of the intermittency route in the presence of noise. By means of comparison, we observe that for a given intensity of noise, which is high enough to transform a subcritical Hopf bifurcation into a continuous one, the secondary bifurcation to a large amplitude limit cycle remains abrupt. We find that a very high value of noise intensity is required for transforming a secondary bifurcation into a continuous transition. We, therefore, conclude that secondary bifurcations can have very high stability due to the presence of higher-order nonlinearities and can appear in turbulent combustion systems despite having relatively high levels of turbulent fluctuations. 

Our study shows that the higher stability of secondary limit cycles reflects the system's high resilience to instability-control strategies, while their abrupt nature makes them exceedingly difficult to predict. Prediction of secondary bifurcation in experiments based on the presented stochastic model will be taken up in future studies.

\section*{Acnowledgements} \label{sec:acknowledgement}
We thank S. Singh, M. Raghunathan, S. A. Pawar, and P. R. Midhun for providing us with the experimental data. We acknowledge J. Dhadphale and A. J. Varghese for their valuable inputs through discussions. R.S.B., A.R., and I.P. are thankful to MHRD for the research assistantship. R.I.S. thanks the Department of Science and Technology for funding this work through the Swarnajayanti and J. C. Bose fellowship (No. JCB/2018/000034/SSC). R.I.S. is also indebted to IIT Madras for its support through the Institute of Eminence (IoE) grant (SB/2021/0845/AE/MHRD/002696).

\appendix

Here, we present the derivation of the slow-flow equations and perform the stochastic averaging to obtain Eqs. (11) and (12) of the manuscript.

\section{Derivation of slow flow amplitude and phase evolution equations}\label{appendix: section1}
Consider the modified stochastic Van der Pol oscillator which is mentioned as Eq. (8) in the main text, describing secondary bifurcation, given as,
\begin{equation}\label{Emodifiedvanderpol2}
\ddot{\eta}+\left ( \mu_6 \eta^6+ \mu_4 \eta^4+\mu_2 \eta^2- \mu_0 \right ) \dot{\eta}+\omega^2 \eta+ \xi=0,
\end{equation}
where $\mu_6$, $\mu_4$, $\mu_2$ are the coefficients of the nonlinear terms and $\mu_0$ is the variable parameter. $\omega$ is the natural frequency of the system. The term $\xi$ is the white noise, with zero mean $\left \langle \xi(t) = 0  \right \rangle$ and with the covariance given as $\psi[\xi,\xi_{\tau}] =\left \langle  \xi(t)\xi(t+\tau) \right \rangle = \Gamma \delta(\tau)$ where $\tau$ and $\Gamma$ are the time of lag and the intensity of the noise, respectively. We assume that white noise is stationary process in our derivation. Using Krylov-Bogoliubov (KB) method of decomposition \cite{krylov2016introduction,balanov2009simple} the general solution for the Eq. (\ref{Emodifiedvanderpol2}) is of the form
\begin{equation}\label{E32}
    \eta(t)= A(t) \cos(\omega(t)t+\phi(t)),
\end{equation}
here, $A(t)$ and $\phi(t)$ are of slow time scale and $\omega(t)$ is of fast time scale. The first derivative $\dot{\eta}$ for the general solution Eq. (\ref{E32}) is given as
\begin{equation}\label{E4.2}
 \dot{\eta}= \dot{A} \cos{(\omega t+\phi)}-A \omega \sin{(\omega t+ \phi )} - A \dot{\phi} \sin{(\omega t+t)}, 
  \end{equation}
 By representing the general solution for $\eta(t)$ in the form of Eq. (\ref{E32}), we introduce two new variables $A(t)$ and $\phi(t)$. In order to remove this ambiguity we specify an additional condition that $A$ and $\phi$ has to satisfy \cite{balanov2009simple} and is given as,
  \begin{equation}\label{E52}
\dot{A} \cos{(\omega t+\phi)} - A \dot{\phi} \sin{(\omega t+t)}=0. 
\end{equation}
Thus, we consider that the general solution has a simple derivative of the form
  \begin{equation}\label{E4.12}
    \dot{\eta} =-A \omega \sin{(\omega t +\phi )}.
\end{equation} 
Writing the general solution in exponential form we get 
 \begin{equation}\label{E62}
\eta=A \cos{(\omega t+\phi)} = A \left ( \frac{e^{i(\omega t+ \phi)}+e^{-i(\omega t+ \phi)}}{2} \right)=\frac{ae^{i\omega t}+a^*e^{-i\omega t}}{2}, 
\end{equation}
where $a=A e^{i \phi}$ and $ a^*=A e^{-i \phi}$. In a similar way we can write Eq. (\ref{E4.12}), Eq. (\ref{E52}) and $\ddot{\eta}$ as
\begin{equation}\label{E82}
    \dot{\eta}= \frac{i\omega (a e^{i \omega t} -a^*e^{-i \omega t})}{2},
\end{equation}

\begin{equation}\label{E72}
    \dot{a} e^{i \omega t}+ \dot{a}^* e^{-i \omega t}=0,
\end{equation}

\begin{equation}\label{E92}
    \ddot{\eta}= i \omega \dot{a} e^{i \omega t}- \frac{\omega^2}{2} (a e^{i \omega t} + a^* e^{-i \omega t}), 
\end{equation}
 respectively, where $\dot{a}=\dot{A} e^{i \phi}+i A \dot{\phi} e^{i \phi}$ and $\dot{a}^*=\dot{A} e^{-i \phi}-i A \dot{\phi} e^{-i \phi}$. Substituting for $\eta$, $\dot{\eta}$ and $\ddot{\eta}$ in Eq. (\ref{Emodifiedvanderpol2}) and letting $a e^{i \omega t}+ a^* e^{-i \omega t}=\beta$ we get
\begin{equation}\label{E102}
 i \omega \dot{a} e^{i \omega t}  - \frac{\omega^2}{2} \beta + \left (\frac{\mu_6}{64} \beta^6 + \frac{\mu_4}{16}\beta^4 + \frac{\mu_2}{4}\beta^2- \mu_0 \right)
    \times  \frac{i \omega}{2} (a e^{i \omega t} - a^* e^{-i \omega t}) + \frac{\omega^2}{2} \beta + \xi= 0,
\end{equation}
here $a$, $\dot{a}$ and $a^*$ are slow functions of time as compared to $e^{(\pm n\omega t)}$, $n$ being an integer. This is expressed as Eq. (10) in the main text. We further simplify Eq. (\ref{E102}) by expanding $\beta^6$, $\beta^4$ and $\beta^2$ using binomial expansion, which is not shown here in the interest of space. In order to eliminate the terms associated with the fast time scale, we divide Eq. (\ref{E102}) with $i\omega e^{i\omega t}$ and  average the whole equation over the time period, $T=2 \pi/\omega$, of fast oscillations. The terms having even integers in $e^{(\pm n \omega t)}$ will be zero after averaging. Substituting for $a=A e^{i \phi}$ , $ a^*=A e^{-i \phi}$ and $aa^* = |A|^2$ we get 
\begin{equation}\label{E112}
    \dot{A}+ i A \dot{\phi} - \frac{\mu_0}{2}A +\frac{\mu_2}{8}A^3 +\frac{\mu_4}{16}A^5 +\frac{5 \mu_6}{128}A^7 -i \frac{\xi}{\omega}e^{-i(\omega t+\phi)}= 0.
\end{equation}
Separating Eq. (\ref{E112}) into real and imaginary parts we have 
\begin{equation}\label{Ereal2}
  \dot{A} - \frac{\mu_0}{2}A +\frac{\mu_2}{8}A^3 +\frac{\mu_4}{16}A^5 +\frac{5 \mu_6}{128}A^7  - \frac{\xi}{\omega}\sin{(\omega t+ \phi)} = 0,
\end{equation}
\begin{equation}\label{Ecomplex2}
    A \dot{\phi}- \frac{\xi}{\omega A}\cos{(\omega t+ \phi)} = 0. 
\end{equation}
Equations (\ref{Ereal2}) and (\ref{Ecomplex2}) are the governing equations for the evolution of slowly varying amplitude and phase, respectively. The amplitude and phase equation can be explicitly written as,
\begin{equation}\label{E7.24}
    \begin{split}
             \dot{A} & = - \left( - \frac{\mu_0}{2}A +\frac{\mu_2}{8}A^3 +\frac{\mu_4}{16}A^5 +\frac{5 \mu_6}{128}A^7  \right)+ \frac{\xi}{\omega}\sin{(\omega t+ \phi)},\\ \dot{\phi} & = \frac{\xi}{\omega A}\cos{(\omega t+ \phi)}
    \end{split}
\end{equation}

\section{Stochastic averaging of the slow flow equations}\label{appendix: section2}
When the stochastic process $\xi = 0$, Eq. (\ref{E7.24}) represents the deterministic evolution of the envelope-amplitude of the limit cycles and phase of the oscillator. In order to simplify the stochastic term $\xi e^{-i\omega t}$ we make use of the procedure used by \citet{stratonovich1963topics}. The method involves the use of Fokker-Planck (FP) equation that describes the time evolution of the joint PDF of amplitude $\mathcal{P}(A)$ and $\mathcal{P}(\phi)$, simplification of the FP equation and reconstructing the stochastic differential equations that correspond to the simplified FP equation. For convenience we write Eq. (\ref{E7.24}) as,
 \begin{equation}
     \begin{split}
         \dot A &= G_A(A,\phi) + H_A(A,\phi,\xi)= F_A, \\
         \dot \phi &= G_{\phi}(A,\phi) + H_{\phi}(A,\phi,\xi)= F_{\phi} 
     \end{split}
 \end{equation}
 where 
 \begin{equation} \label{7.25}
     \begin{split}
         &G_A =  - \left(- \frac{\mu_0}{2}A +\frac{\mu_2}{8}A^3 +\frac{\mu_4}{16}A^5 +\frac{5 \mu_6}{128}A^7 \right), \\ &H_A = \frac{\xi}{\omega}\sin{(\omega t+ \phi)}, \\ &G_\phi= 0, \\ &H_\phi= \frac{\xi}{\omega A}\cos{(\omega t+ \phi)}.
     \end{split}
 \end{equation}
 $F_A$ and $F_\phi$ are the stochastic functions of amplitude and phase. Following \citet{stratonovich1963topics} and \citet{balanov2009simple}, we write the FP equation describing the joint probability density function $\mathcal{P}(A,\phi,t)$ as,
 \begin{equation} \label{E7.27}
     \begin{split}
         \frac{\partial \mathcal P}{\partial t} = &- \frac{\partial}{\partial A} \left \{ \left ( \left \langle F_A  \right \rangle + \int_{t_0-t}^{0} \psi \left [ \frac{\partial F_A}{\partial A}, F_{A\tau} \right ]d\tau + \int_{t_0-t}^{0} \psi \left [ \frac{\partial F_A}{\partial \phi}, F_{\phi\tau} \right ]d\tau \right) \mathcal{P} \right \} \\
         &-\frac{\partial}{\partial \phi} \left \{ \left ( \left \langle F_\phi  \right \rangle + \int_{t_0-t}^{0} \psi \left [ \frac{\partial F_\phi}{\partial A}, F_{A\tau} \right ]d\tau + \int_{t_0-t}^{0} \psi \left [ \frac{\partial F_\phi}{\partial \phi}, F_{\phi\tau} \right ]d\tau \right) \mathcal{P} \right \} \\
         &+ \frac{\partial^2}{\partial A^2} \left \{ \left ( \int_{t_0-t}^{0} \psi \left [ F_A, F_{A\tau} \right ] d\tau \right) \mathcal{P} \right \} +\frac{\partial^2}{\partial A \partial \phi} \left \{ \left ( \int_{t_0-t}^{0} \psi \left [ F_A, F_{\phi\tau} \right ] d\tau \right) \mathcal{P} \right \} \\
         &+\frac{\partial^2}{\partial \phi \partial A} \left \{ \left ( \int_{t_0-t}^{0} \psi \left [ F_\phi, F_{A\tau} \right ] d\tau \right) \mathcal{P} \right \}
         +\frac{\partial^2}{\partial \phi^2} \left \{ \left ( \int_{t_0-t}^{0} \psi \left [ F_\phi, F_{\phi\tau} \right ] d\tau \right) \mathcal{P} \right \}
     \end{split}
 \end{equation}
   Here, $\psi \left [ X, Y_\tau \right ]$ is the cross-covarience of the two stochastic process $X$ and $Y_t$ at time instants $t$ and $t+\tau$, respectively. The symbol $\left \langle . \right \rangle$ represents the ensemble of realizations of the stochastic process. The covariance terms that appear in Eq. (\ref{E7.27}) can be simplified as follows. To begin with, we consider the first covariance term 
   \begin{equation} \label{E7.28}
       \begin{split}
           \psi \left [ \frac{\partial F_A}{\partial A}, F_{A\tau} \right ] &=
           \left \langle \frac{\partial F_A}{\partial A} \times F_{A\tau} \right \rangle -  \left \langle \frac{\partial F_A}{\partial A} \right \rangle \left \langle  F_{A\tau} \right \rangle \\ &= \left \langle \frac{\partial(G_A + H_A)}{\partial A} \times (G_{A\tau}+H_{A\tau}) \right \rangle \\ &- \left \langle \frac{\partial (G_A + H_A)}{\partial A} \right \rangle \left \langle G_{A\tau} + H_{A\tau} \right \rangle 
       \end{split}
   \end{equation}
$G_A$ and $\partial G_A / \partial A$ are deterministic functions of time and they remain same for any realization of stochastic process $\xi$. Hence, their ensemble averages are given by
\begin{equation} \label{E7.29}
     \left \langle G_A \right \rangle = G_A, \hspace{10mm} \left \langle \frac{\partial G_A}{\partial A}  \right \rangle =  \frac{\partial G_A}{\partial A}.
\end{equation}
Considering Eq. (\ref{E7.29}) and the fact, that the ensemble average of a product of a deterministic and a stochastic functions can be written as the product of deterministic function and the average of the stochastic function. Eq. (\ref{E7.28}) can be written as 
\begin{equation}\label{7.31}
    \psi \left [ \frac{\partial F_A}{\partial A}, F_{A\tau} \right ] = \left \langle \frac{\partial H_A}{\partial A} \times H_{A\tau} \right \rangle = \left \langle 0 \times H_{A\tau} \right \rangle = 0
\end{equation}
We can also calculate the the averages and covariances of other terms in Eq. (\ref{E7.27}) as
\begin{equation} \label{E7.35.0}
\begin{split}
         \psi \left [ \frac{\partial F_A}{\partial \phi}, F_{\phi\tau} \right ] &= \left \langle \frac{\partial H_A}{\partial \phi} \times H_{\phi\tau} \right \rangle \\ &= \left \langle \frac{\xi}{\omega} \cos \left (\omega t+ \phi \right) \times \left( \frac{\xi_\tau}{\omega A_\tau} \cos \left( \omega t + \omega \tau + \phi_\tau \right) \right )  \right \rangle \\ &=\left \langle \xi\xi_\tau \right \rangle \frac{1}{ A_\tau \omega^2} \cos \left (\omega t+ \phi \right)  \cos \left( \omega t + \omega \tau + \phi_\tau \right) 
\end{split}
\end{equation}
Now, we evaluate an integral of $\psi [ \partial F_A/\partial \phi, F_{\phi\tau}]$ over $\tau$ from $(t_0 - t)$ to 0 where $t_0$ is some initial time moment from which we start to consider the process. We set $t_0$ to minus infinity so as to consider an established process. We also make an assumption that noise $\xi$ is a fast stochastic process whose correlation time is much lesser than the systems relaxation time \cite{balanov2009simple}. Hence the slow variables can be treated as constant in the time interval which is of the order of the correlation time of the noise, which implies $A_\tau=A$ and $\phi_\tau=\phi$. After simplifying the trigonometric terms in Eq. (\ref{E7.35.0}) we can write the integral as
\begin{equation} \label{E7.37}
    \begin{split}
        \int_{-\infty}^0 \psi \left [ \frac{\partial F_A}{\partial \phi}, F_{\phi\tau} \right ] d\tau &= \frac{1+\cos (2\omega t + 2 \phi)}{2 A \omega^2} \int_{-\infty}^0 \left \langle \xi \xi_\tau \right \rangle \cos (\omega \tau) d\tau \\ &-\frac{\sin (2\omega t + 2 \phi)}{2 A \omega^2} \int_{-\infty}^0 \left \langle \xi \xi_\tau \right \rangle \sin (\omega \tau) d\tau 
    \end{split}
\end{equation}
We have initially assumed that the noise we are considering is a stationary process, then its correlation function $\left \langle  \xi \xi_\tau \right \rangle$ depends only on $\tau$. According to Wiener-Khintchine theorem \cite{coffey2012langevin}, the autocorrelation of the stationary process is the Fourier transform of the power spectral density $\Gamma$.  The first integral on the right hand side of the Eq. (\ref{E7.37}) is half of the Fourier transform (FT) of the correlation function  $\left \langle  \xi \xi_\tau \right \rangle$ which is equal to $\Gamma/2$. The second integral is the imaginary part of the FT and is equal to zero. Hence Eq. (\ref{E7.37}) simplifies to 

\begin{equation} \label{E7.38}
    \begin{split}
        \int_{-\infty}^0 \psi \left [ \frac{\partial F_A}{\partial \phi}, F_{\phi\tau} \right ] d\tau &= \frac{\Gamma}{4 A \omega^2} (1+\cos (2\omega t + 2 \phi))
    \end{split}
\end{equation}
We can again apply the Krylo-Bogoliubov method of averaging to Eq. (\ref{E7.38}) by taking $A$ and $\phi$ as slowly varying functions of time to obtain
\begin{equation}\label{E7.39}
     \int_{-\infty}^0 \psi \left [ \frac{\partial F_A}{\partial \phi}, F_{\phi\tau} \right ] d\tau = \frac{\Gamma}{4 A \omega^2}
\end{equation}
In a similar manner we can simplify the other terms of the Eq. (\ref{E7.27}) as
\begin{equation}\label{E7.40}
    \psi \left [ \frac{\partial F_\phi}{\partial A}, F_{A\tau} \right ] =  \left \langle \frac{\partial H_\phi}{\partial A} \times H_{A\tau} \right \rangle = 0
\end{equation}
\begin{equation}\label{E7.41}
    \psi \left [ \frac{\partial F_\phi}{\partial \phi}, F_{\phi\tau} \right ]=\left \langle \frac{\partial H_\phi}{\partial \phi} \times H_{\phi\tau} \right \rangle = 0
\end{equation}
\begin{equation}\label{E7.42}
    \psi \left [ F_A, F_{A\tau} \right]= \left \langle H_A H_{A\tau} \right \rangle =\frac{\Gamma}{4 \omega^2}
\end{equation}
\begin{equation}\label{E7.43}
    \psi \left [ F_A, F_{\phi\tau} \right]= \left \langle H_A H_{\phi\tau} \right \rangle=0
\end{equation}
\begin{equation}\label{E7.44}
      \psi \left [ F_\phi, F_{A\tau} \right]= \left \langle H_\phi H_{A\tau} \right \rangle=0
\end{equation}
\begin{equation}\label{E7.45}
    \psi \left [ F_\phi, F_{\phi\tau} \right]= \left \langle H_\phi H_{\phi\tau} \right \rangle = \frac{\Gamma}{4 \omega^2 A^2}
\end{equation}
In the view of the above, Eq. (\ref{E7.27}) can be rewritten as
\begin{equation} \label{E7.46}
\begin{split}
    \frac{\partial \mathcal P}{\partial t} &= \frac{\partial}{\partial A} \left \{ \left ( G_A  + \int_{-\infty}^{0} \psi \left  \langle \frac{\partial H_A}{\partial A}, H_{A\tau} \right \rangle d\tau + \int_{-\infty}^{0} \psi \left  \langle \frac{\partial H_A}{\partial \phi}, H_{\phi\tau} \right \rangle d\tau \right) \mathcal{P} \right \} \\
    &-\frac{\partial}{\partial \phi} \left \{ \left ( G_\phi + \int_{-\infty}^{0} \psi \left  \langle \frac{\partial H_\phi}{\partial A}, H_{A\tau} \right \rangle d\tau + \int_{-\infty}^{0} \psi \left  \langle \frac{\partial H_\phi}{\partial \phi}, H_{\phi\tau} \right \rangle d\tau \right) \mathcal{P} \right \} \\
     &+ \frac{\partial^2}{\partial A^2} \left \{ \left ( \int_{-\infty}^{0} \psi \left  \langle H_A, H_{A\tau} \right \rangle  d\tau \right) \mathcal{P} \right \} +\frac{\partial^2}{\partial A \partial \phi} \left \{ \left ( \int_{-\infty}^{0} \psi \left  \langle H_A, H_{\phi\tau} \right \rangle  d\tau \right) \mathcal{P} \right \} \\
     &+\frac{\partial^2}{\partial \phi \partial A} \left \{ \left ( \int_{-\infty}^{0} \psi \left  \langle H_\phi, H_{A\tau} \right \rangle  d\tau \right) \mathcal{P} \right \}
     +\frac{\partial^2}{\partial \phi^2} \left \{ \left ( \int_{-\infty}^{0} \psi \left  \langle H_\phi, H_{\phi\tau} \right \rangle  d\tau \right) \mathcal{P} \right \}
\end{split}
\end{equation}
Substituting all the terms in Eq. (\ref{E7.39})-(\ref{E7.45}) into Eq. (\ref{E7.46}) we obtain
\begin{equation}\label{E7.47}
    \begin{split}
        \frac{\partial \mathcal P}{\partial t} = &- \frac{\partial}{\partial A} \left \{ \left ( G_A + \frac{\Gamma}{4A \omega^2} \right) \mathcal{P} \right \}- \frac{\partial}{\partial \phi}\{G_\phi \mathcal{P}\} \\ 
        &+ \frac{\partial^2}{\partial A^2} \left \{ \frac{\Gamma}{4 \omega^2} \mathcal{P} \right \} + \frac{\partial^2}{\partial \phi^2} \left \{ \frac{\Gamma}{4 \omega^2 A^2} \mathcal{P} \right \}.
    \end{split}
\end{equation}
Equation (\ref{E7.47}) is a Fokker-Planck equation which is simplified by means of averaging over the period of fast time scale $T= 2\pi/\omega$. Now, we would like reconstruct stochastic equations in the form
\begin{equation}\label{E7.48}
    \dot A = \Tilde{G_A}(A,\phi) + \Tilde{H_A}(A,\phi, \xi_1),
\end{equation} 
\begin{equation}\label{E7.49}
    \dot \phi = \Tilde{G_\phi}(A,\phi) + \Tilde{H_\phi}(A,\phi, \xi_2),
\end{equation}
that would result in the simplified FP Eq. (\ref{E7.47}), if one wanted to construct it following the Eq. (\ref{E7.27}). We find the expressions for $\Tilde{G_A}, \Tilde{H_A}, \Tilde{G_\phi}$ and $\Tilde{H_\phi}$ by comparing seperate terms of Eq. (\ref{E7.47}) with the corresponding terms of Eq. (\ref{E7.46}), considering that all the functions in the latter would be marked by tildes. We observe that 
\begin{equation}\label{E7.50.0}
    \int_{-\infty}^0 \left \langle \Tilde{H_A} \Tilde{H_{\phi \tau}} \right \rangle d\tau = \int_{-\infty}^0 \left \langle \Tilde{H_\phi} \Tilde{H_{A \tau}} \right \rangle d\tau = 0,
\end{equation}
This can be true if the process $\Tilde{H_A}$ and $\Tilde{H_\phi}$ are not correlated. If the Eq. (\ref{E7.50.0}) is true, then the two pairs of processes  $\partial \Tilde{H_A}/\partial \phi$ and $\Tilde{H_\phi}$, and $\partial \Tilde{H_\phi}/\partial A$ and $\Tilde{H_A}$ are not correlated that is
\begin{equation}\label{E7.50}
    \int_{-\infty}^0 \left \langle \frac{\partial \Tilde{H_A}}{\partial \phi}  \times \Tilde{H_{\phi \tau}}  \right \rangle d\tau = \int_{-\infty}^0 \left \langle \frac{\partial \Tilde{H_\phi}}{\partial A} \times \Tilde{H_{A\tau}}  \right \rangle d\tau=0.
\end{equation}
Therefore we have
\begin{equation}\label{E7.51}
    \Tilde{G_A}  + \int_{-\infty}^{0} \psi \left  \langle \frac{\partial \Tilde{H_A}}{\partial A} \times \Tilde{H_{A\tau}} \right \rangle d\tau = G_A + \frac{\Gamma}{4 A \omega^2},
\end{equation}

\begin{equation}\label{E7.52}
    \Tilde{G_\phi}  + \int_{-\infty}^{0} \psi \left  \langle \frac{\partial \Tilde{H_{\phi}}}{\partial \phi} \times \Tilde{H_{\phi\tau}} \right \rangle d\tau = G_\phi.
\end{equation}
Next we consider
\begin{equation}\label{E7.53}
   \int_{-\infty}^{0} \psi \left  \langle \Tilde{H_A} \times \Tilde{H_{A\tau}} \right \rangle d\tau = \frac{\Gamma}{4 \omega^2},
\end{equation}
which is an expression independent of $A$, and therefore $\partial \Tilde{H_A}/\partial A = 0$. This leads to the disappearance of the integral in Eq. (\ref{E7.51}), and the final expression for $\Tilde{G_A}$ is 
\begin{equation}
    \Tilde{G_A} = G_A + \frac{\Gamma}{4 A \omega^2}.
\end{equation}
Next, we consider
\begin{equation}\label{E7.55}
   \int_{-\infty}^{0} \psi \left  \langle \Tilde{H_\phi} \times \Tilde{H_{\phi\tau}} \right \rangle d\tau = \frac{\Gamma}{4 \omega^2 A^2},
\end{equation}
Here, the integral depends on $A$, but does not depend on $\phi$, therefore the term involving $\partial \title{H_\phi}/\partial \phi$ vanishes, and $\Tilde{G_\phi}$ is given as
\begin{equation}
    \Tilde{G_\phi}= G_\phi.
\end{equation}
Equation (\ref{E7.53}) and (\ref{E7.55}) are only valid if $\Tilde{H_A}$ and $\Tilde{H_\phi}$ can be expressed as 
\begin{equation}
    \begin{split}
        \Tilde{H_A}= \frac{\sqrt{\Gamma}}{\sqrt{2}\omega}\xi_1 \hspace{10mm} \mathrm{and} \hspace{10mm} \Tilde{H_\phi}= \frac{\sqrt{\Gamma}}{\sqrt{2}\omega A}\xi_2 
    \end{split}
\end{equation}
Where $\xi_1$ and $\xi_2$ are delta-correlated noise with zero mean and unity variance. In order for $\Tilde{H_A}$ and $\Tilde{H_\phi}$ to be uncorrelated, we need that $\xi_1$ 
and $\xi_2$ are uncorrelated which is given as
\begin{equation}
    \left \langle \xi_1(t) \xi_2(t+\tau) \right \rangle =0.
\end{equation}
Finally, we can write the simplified stochastic differential Eq. (\ref{E7.24}) as 
\begin{equation}\label{E7.58}
    \begin{split}
             \dot{A} & = - \left( - \frac{\mu_0}{2}A +\frac{\mu_2}{8}A^3 +\frac{\mu_4}{16}A^5 +\frac{5 \mu_6}{128}A^7 \right)+ \frac{\Gamma}{4A\omega^2} + \frac{\sqrt{\Gamma}}{\sqrt{2}\omega}\xi_1,\\ \dot{\phi} & = \frac{\sqrt{\Gamma}}{\sqrt{2}\omega A}\xi_2.
    \end{split}
\end{equation}
This is expressed in Eq. (11) and Eq. (12) of the main text.
 \bibliographystyle{elsarticle-num-names} 
 \bibliography{manu}

\begin{thebibliography}{54}
\expandafter\ifx\csname natexlab\endcsname\relax\def\natexlab#1{#1}\fi
\providecommand{\url}[1]{\texttt{#1}}
\providecommand{\href}[2]{#2}
\providecommand{\path}[1]{#1}
\providecommand{\DOIprefix}{doi:}
\providecommand{\ArXivprefix}{arXiv:}
\providecommand{\URLprefix}{URL: }
\providecommand{\Pubmedprefix}{pmid:}
\providecommand{\doi}[1]{\href{http://dx.doi.org/#1}{\path{#1}}}
\providecommand{\Pubmed}[1]{\href{pmid:#1}{\path{#1}}}
\providecommand{\bibinfo}[2]{#2}
\ifx\xfnm\relax \def\xfnm[#1]{\unskip,\space#1}\fi
\bibitem[{Lieuwen and Yang(2005)}]{lieuwen2005combustion}
\bibinfo{author}{T.~C. Lieuwen}, \bibinfo{author}{V.~Yang},
  \bibinfo{title}{Combustion Instabilities in Gas Turbine Engines: Operational
  Experience, Fundamental Mechanisms, and Modeling}, Progress in Aeronautics
  and Astronautics, \bibinfo{publisher}{American Institute of Aeronautics and
  Astronautics}, \bibinfo{year}{2005}. \URLprefix
  \url{https://doi.org/10.2514/4.866807}.
\bibitem[{Juniper and Sujith(2018)}]{juniper2018sensitivity}
\bibinfo{author}{M.~P. Juniper}, \bibinfo{author}{R.~I. Sujith},
\newblock \bibinfo{title}{Sensitivity and nonlinearity of thermoacoustic
  oscillations},
\newblock \bibinfo{journal}{Annu. Rev. Fluid Mech.} \bibinfo{volume}{50}
  (\bibinfo{year}{2018}) \bibinfo{pages}{661--689}. \URLprefix
  \url{https://doi.org/10.1146/annurev-fluid-122316-045125}.
\bibitem[{Sujith and Pawar(2021)}]{sujith2021thermoacoustic}
\bibinfo{author}{R.~I. Sujith}, \bibinfo{author}{S.~Pawar},
  \bibinfo{title}{Thermoacoustic Instability: A Complex Systems Perspective},
  Springer Series in Synergetics, \bibinfo{publisher}{Springer International
  Publishing}, \bibinfo{year}{2021}. \URLprefix
  \url{https://doi.org/10.1007/978-3-030-81135-8}.
\bibitem[{Rayleigh(1878)}]{rayleigh1878explanation}
\bibinfo{author}{L.~Rayleigh},
\newblock \bibinfo{title}{The explanation of certain acoustical phenomena},
\newblock \bibinfo{journal}{Roy. Inst. Proc.} \bibinfo{volume}{8}
  (\bibinfo{year}{1878}) \bibinfo{pages}{536--542}. \URLprefix
  \url{https://doi.org/10.1038/018319a0}.
\bibitem[{Chu(1965)}]{chu1965energy}
\bibinfo{author}{B.-T. Chu},
\newblock \bibinfo{title}{On the energy transfer to small disturbances in fluid
  flow (part {I})},
\newblock \bibinfo{journal}{Acta Mech.} \bibinfo{volume}{1}
  (\bibinfo{year}{1965}) \bibinfo{pages}{215--234}. \URLprefix
  \url{https://doi.org/10.1007/BF01387235}.
\bibitem[{Strogatz(2018)}]{strogatz2018nonlinear}
\bibinfo{author}{S.~H. Strogatz}, \bibinfo{title}{Nonlinear Dynamics and Chaos,
  With Applications to Physics, Biology, Chemistry, and Engineering},
  \bibinfo{publisher}{CRC press}, \bibinfo{year}{2018}. \URLprefix
  \url{https://doi.org/10.1201/9780429492563}.
\bibitem[{Lieuwen(2002)}]{lieuwen2002experimental}
\bibinfo{author}{T.~C. Lieuwen},
\newblock \bibinfo{title}{Experimental investigation of limit-cycle
  oscillations in an unstable gas turbine combustor},
\newblock \bibinfo{journal}{J. Propuls. Power} \bibinfo{volume}{18}
  (\bibinfo{year}{2002}) \bibinfo{pages}{61--67}. \URLprefix
  \url{https://doi.org/10.2514/2.5898}.
\bibitem[{Moeck et~al.(2008)Moeck, Bothien, Schimek, Lacarelle, and
  Paschereit}]{moeck2008subcritical}
\bibinfo{author}{J.~Moeck}, \bibinfo{author}{M.~Bothien},
  \bibinfo{author}{S.~Schimek}, \bibinfo{author}{A.~Lacarelle},
  \bibinfo{author}{C.~Paschereit},
\newblock \bibinfo{title}{Subcritical thermoacoustic instabilities in a
  premixed combustor},
\newblock in: \bibinfo{booktitle}{14th AIAA/CEAS aeroacoustics conference (29th
  AIAA aeroacoustics conference)}, \bibinfo{year}{2008}, pp.
  \bibinfo{pages}{2008--2946}. \URLprefix
  \url{https://doi.org/10.2514/6.2008-2946}.
\bibitem[{Li et~al.(2017)Li, Zhao, and Yang}]{li2017experimental}
\bibinfo{author}{X.~Li}, \bibinfo{author}{D.~Zhao}, \bibinfo{author}{X.~Yang},
\newblock \bibinfo{title}{Experimental and theoretical bifurcation study of a
  nonlinear standing-wave thermoacoustic system},
\newblock \bibinfo{journal}{Energy} \bibinfo{volume}{135}
  (\bibinfo{year}{2017}) \bibinfo{pages}{553--562}. \URLprefix
  \url{https://doi.org/10.1016/j.energy.2017.06.134}.
\bibitem[{Juniper(2012)}]{juniper2012triggering}
\bibinfo{author}{M.~P. Juniper},
\newblock \bibinfo{title}{Triggering in thermoacoustics},
\newblock \bibinfo{journal}{Int. J. Spray Combust. Dyn.} \bibinfo{volume}{4}
  (\bibinfo{year}{2012}) \bibinfo{pages}{217--237}. \URLprefix
  \url{https://doi.org/10.1260%2F1756-8277.4.3.217}.
\bibitem[{Etikyala and Sujith(2017)}]{etikyala2017change}
\bibinfo{author}{S.~Etikyala}, \bibinfo{author}{R.~I. Sujith},
\newblock \bibinfo{title}{Change of criticality in a prototypical
  thermoacoustic system},
\newblock \bibinfo{journal}{Chaos} \bibinfo{volume}{27} (\bibinfo{year}{2017})
  \bibinfo{pages}{023106}. \URLprefix \url{https://doi.org/10.1063/1.4975822}.
\bibitem[{Subramanian et~al.(2010)Subramanian, Mariappan, Sujith, and
  Wahi}]{subramanian2010bifurcation}
\bibinfo{author}{P.~Subramanian}, \bibinfo{author}{S.~Mariappan},
  \bibinfo{author}{R.~I. Sujith}, \bibinfo{author}{P.~Wahi},
\newblock \bibinfo{title}{Bifurcation analysis of thermoacoustic instability in
  a horizontal {R}ijke tube},
\newblock \bibinfo{journal}{Int. J. Spray Combust. Dyn.} \bibinfo{volume}{2}
  (\bibinfo{year}{2010}) \bibinfo{pages}{325--355}. \URLprefix
  \url{https://doi.org/10.1260%2F1756-8277.2.4.325}.
\bibitem[{Ananthkrishnan et~al.(1998)Ananthkrishnan, Sudhakar, Sudershan, and
  Agarwal}]{ananthkrishnan1998application}
\bibinfo{author}{N.~Ananthkrishnan}, \bibinfo{author}{K.~Sudhakar},
  \bibinfo{author}{S.~Sudershan}, \bibinfo{author}{A.~Agarwal},
\newblock \bibinfo{title}{Application of secondary bifurcations to
  large-amplitude limit cycles in mechanical systems},
\newblock \bibinfo{journal}{J. Sound Vib.} \bibinfo{volume}{215}
  (\bibinfo{year}{1998}) \bibinfo{pages}{183--188}. \URLprefix
  \url{https://doi.org/10.1006/jsvi.1998.1623}.
\bibitem[{Ananthkrishnan et~al.(2005)Ananthkrishnan, Deo, and
  Culick}]{ananthkrishnan2005reduced}
\bibinfo{author}{N.~Ananthkrishnan}, \bibinfo{author}{S.~Deo},
  \bibinfo{author}{F.~E. Culick},
\newblock \bibinfo{title}{Reduced-order modeling and dynamics of nonlinear
  acoustic waves in a combustion chamber},
\newblock \bibinfo{journal}{Combust. Sci. and Tech.} \bibinfo{volume}{177}
  (\bibinfo{year}{2005}) \bibinfo{pages}{221--248}. \URLprefix
  \url{https://doi.org/10.1080/00102200590900219}.
\bibitem[{Nalini et~al.(2015)Nalini, Maria, Alessandra, Vishnu, Samadhan, and
  Sujith}]{nalininonlinear}
\bibinfo{author}{M.~Nalini}, \bibinfo{author}{H.~Maria},
  \bibinfo{author}{B.~Alessandra}, \bibinfo{author}{R.~Vishnu},
  \bibinfo{author}{P.~Samadhan}, \bibinfo{author}{R.~I. Sujith},
\newblock \bibinfo{title}{Nonlinear dynamics of a laminar {V}-flame in a
  combustor},
\newblock in: \bibinfo{booktitle}{22nd Int. Congr. Sound Vib. ICSV},
  \bibinfo{year}{2015}, pp. \bibinfo{pages}{1--8}.
\bibitem[{Roy et~al.(2021)Roy, Singh, Nair, Chaudhuri, and
  Sujith}]{roy2021flame}
\bibinfo{author}{A.~Roy}, \bibinfo{author}{S.~Singh},
  \bibinfo{author}{A.~Nair}, \bibinfo{author}{S.~Chaudhuri},
  \bibinfo{author}{R.~I. Sujith},
\newblock \bibinfo{title}{Flame dynamics during intermittency and secondary
  bifurcation to longitudinal thermoacoustic instability in a swirl-stabilized
  annular combustor},
\newblock \bibinfo{journal}{Proc. Combust. Inst.} \bibinfo{volume}{38}
  (\bibinfo{year}{2021}) \bibinfo{pages}{6221--6230}. \URLprefix
  \url{https://doi.org/10.1016/j.proci.2020.08.053}.
\bibitem[{Singh et~al.(2021)Singh, Roy, Reeja, Nair, Chaudhuri, and
  Sujith}]{singh2021intermittency}
\bibinfo{author}{S.~Singh}, \bibinfo{author}{A.~Roy}, \bibinfo{author}{K.~V.
  Reeja}, \bibinfo{author}{A.~Nair}, \bibinfo{author}{S.~Chaudhuri},
  \bibinfo{author}{R.~I. Sujith},
\newblock \bibinfo{title}{Intermittency, secondary bifurcation and mixed-mode
  oscillations in a swirl-stabilized annular combustor: Experiments and
  modeling},
\newblock \bibinfo{journal}{J. Eng. Gas Turbines Power.} \bibinfo{volume}{143}
  (\bibinfo{year}{2021}) \bibinfo{pages}{051028}. \URLprefix
  \url{https://doi.org/10.1115/1.4049407}.
\bibitem[{Wang et~al.(2021)Wang, Han, Song, Yang, and Sung}]{wang2021multi}
\bibinfo{author}{X.~Wang}, \bibinfo{author}{X.~Han}, \bibinfo{author}{H.~Song},
  \bibinfo{author}{D.~Yang}, \bibinfo{author}{C.-J. Sung},
\newblock \bibinfo{title}{Multi-bifurcation behaviors of stability regimes in a
  centrally staged swirl burner},
\newblock \bibinfo{journal}{Phys. Fluids} \bibinfo{volume}{33}
  (\bibinfo{year}{2021}) \bibinfo{pages}{095121}. \URLprefix
  \url{https://doi.org/10.1063/5.0063562}.
\bibitem[{Candel et~al.(2009)Candel, Durox, Ducruix, Birbaud, Noiray, and
  Schuller}]{candel2009flame}
\bibinfo{author}{S.~Candel}, \bibinfo{author}{D.~Durox},
  \bibinfo{author}{S.~Ducruix}, \bibinfo{author}{A.-L. Birbaud},
  \bibinfo{author}{N.~Noiray}, \bibinfo{author}{T.~Schuller},
\newblock \bibinfo{title}{Flame dynamics and combustion noise: progress and
  challenges},
\newblock \bibinfo{journal}{Int. J. Aeroacoustics} \bibinfo{volume}{8}
  (\bibinfo{year}{2009}) \bibinfo{pages}{1--56}. \URLprefix
  \url{https://doi.org/10.1260%2F147547209786234984}.
\bibitem[{Gotoda et~al.(2011)Gotoda, Nikimoto, Miyano, and
  Tachibana}]{gotoda2011dynamic}
\bibinfo{author}{H.~Gotoda}, \bibinfo{author}{H.~Nikimoto},
  \bibinfo{author}{T.~Miyano}, \bibinfo{author}{S.~Tachibana},
\newblock \bibinfo{title}{Dynamic properties of combustion instability in a
  lean premixed gas-turbine combustor},
\newblock \bibinfo{journal}{Chaos} \bibinfo{volume}{21} (\bibinfo{year}{2011})
  \bibinfo{pages}{013124}. \URLprefix \url{https://doi.org/10.1063/1.3563577}.
\bibitem[{Nair et~al.(2013)Nair, Thampi, Karuppusamy, Gopalan, and
  Sujith}]{nair2013loss}
\bibinfo{author}{V.~Nair}, \bibinfo{author}{G.~Thampi},
  \bibinfo{author}{S.~Karuppusamy}, \bibinfo{author}{S.~Gopalan},
  \bibinfo{author}{R.~I. Sujith},
\newblock \bibinfo{title}{Loss of chaos in combustion noise as a precursor of
  impending combustion instability},
\newblock \bibinfo{journal}{Int. J. Spray Combust. Dyn.} \bibinfo{volume}{5}
  (\bibinfo{year}{2013}) \bibinfo{pages}{273--290}. \URLprefix
  \url{https://doi.org/10.1260%2F1756-8277.5.4.273}.
\bibitem[{Tony et~al.(2015)Tony, Gopalakrishnan, Sreelekha, and
  Sujith}]{tony2015detecting}
\bibinfo{author}{J.~Tony}, \bibinfo{author}{E.~Gopalakrishnan},
  \bibinfo{author}{E.~Sreelekha}, \bibinfo{author}{R.~I. Sujith},
\newblock \bibinfo{title}{Detecting deterministic nature of pressure
  measurements from a turbulent combustor},
\newblock \bibinfo{journal}{Phys. Rev. E} \bibinfo{volume}{92}
  (\bibinfo{year}{2015}) \bibinfo{pages}{062902}. \URLprefix
  \url{https://doi.org/10.1103/PhysRevE.92.062902}.
\bibitem[{Nair and Sujith(2014)}]{nair2014multifractality}
\bibinfo{author}{V.~Nair}, \bibinfo{author}{R.~I. Sujith},
\newblock \bibinfo{title}{Multifractality in combustion noise: predicting an
  impending combustion instability},
\newblock \bibinfo{journal}{J. Fluid Mech.} \bibinfo{volume}{747}
  (\bibinfo{year}{2014}) \bibinfo{pages}{635--655}. \URLprefix
  \url{https://doi.org/10.1017/jfm.2014.171}.
\bibitem[{Nair et~al.(2014)Nair, Thampi, and Sujith}]{nair2014intermittency}
\bibinfo{author}{V.~Nair}, \bibinfo{author}{G.~Thampi}, \bibinfo{author}{R.~I.
  Sujith},
\newblock \bibinfo{title}{Intermittency route to thermoacoustic instability in
  turbulent combustors},
\newblock \bibinfo{journal}{J. Fluid Mech.} \bibinfo{volume}{756}
  (\bibinfo{year}{2014}) \bibinfo{pages}{470--487}. \URLprefix
  \url{https://doi.org/10.1017/jfm.2014.468}.
\bibitem[{Gotoda et~al.(2014)Gotoda, Shinoda, Kobayashi, Okuno, and
  Tachibana}]{gotoda2014detection}
\bibinfo{author}{H.~Gotoda}, \bibinfo{author}{Y.~Shinoda},
  \bibinfo{author}{M.~Kobayashi}, \bibinfo{author}{Y.~Okuno},
  \bibinfo{author}{S.~Tachibana},
\newblock \bibinfo{title}{Detection and control of combustion instability based
  on the concept of dynamical system theory},
\newblock \bibinfo{journal}{Phys. Rev. E} \bibinfo{volume}{89}
  (\bibinfo{year}{2014}) \bibinfo{pages}{022910}. \URLprefix
  \url{https://doi.org/10.1103/PhysRevE.89.022910}.
\bibitem[{Huang(2015)}]{huang2015advanced}
\bibinfo{author}{Y.~Huang}, \bibinfo{title}{Advanced methods for validating
  combustion instability predictions using pressure measurements}, Ph.D.
  thesis, Purdue University, \bibinfo{year}{2015}.
\bibitem[{Kabiraj et~al.(2015)Kabiraj, Saurabh, Karimi, Sailor, Mastorakos,
  Dowling, and Paschereit}]{kabiraj2015chaos}
\bibinfo{author}{L.~Kabiraj}, \bibinfo{author}{A.~Saurabh},
  \bibinfo{author}{N.~Karimi}, \bibinfo{author}{A.~Sailor},
  \bibinfo{author}{E.~Mastorakos}, \bibinfo{author}{A.~P. Dowling},
  \bibinfo{author}{C.~O. Paschereit},
\newblock \bibinfo{title}{Chaos in an imperfectly premixed model combustor},
\newblock \bibinfo{journal}{Chaos} \bibinfo{volume}{25} (\bibinfo{year}{2015})
  \bibinfo{pages}{023101}. \URLprefix \url{https://doi.org/10.1063/1.4906943}.
\bibitem[{Kheirkhah et~al.(2017)Kheirkhah, Cirtwill, Saini, Venkatesan, and
  Steinberg}]{kheirkhah2017dynamics}
\bibinfo{author}{S.~Kheirkhah}, \bibinfo{author}{J.~M. Cirtwill},
  \bibinfo{author}{P.~Saini}, \bibinfo{author}{K.~Venkatesan},
  \bibinfo{author}{A.~M. Steinberg},
\newblock \bibinfo{title}{Dynamics and mechanisms of pressure, heat release
  rate, and fuel spray coupling during intermittent thermoacoustic oscillations
  in a model aeronautical combustor at elevated pressure},
\newblock \bibinfo{journal}{Combust. Flame} \bibinfo{volume}{185}
  (\bibinfo{year}{2017}) \bibinfo{pages}{319--334}. \URLprefix
  \url{https://doi.org/10.1016/j.combustflame.2017.07.017}.
\bibitem[{Clavin et~al.(1994)Clavin, Kim, and Williams}]{clavin1994turbulence}
\bibinfo{author}{P.~Clavin}, \bibinfo{author}{J.~Kim},
  \bibinfo{author}{F.~Williams},
\newblock \bibinfo{title}{Turbulence-induced noise effects on high-frequency
  combustion instabilities},
\newblock \bibinfo{journal}{Combust. Sci. and Tech.} \bibinfo{volume}{96}
  (\bibinfo{year}{1994}) \bibinfo{pages}{61--84}. \URLprefix
  \url{https://doi.org/10.1080/00102209408935347}.
\bibitem[{Burnley and Culick(2000)}]{burnley2000influence}
\bibinfo{author}{V.~Burnley}, \bibinfo{author}{F.~Culick},
\newblock \bibinfo{title}{Influence of random excitations on acoustic
  instabilities in combustion chambers},
\newblock \bibinfo{journal}{AIAA J.} \bibinfo{volume}{38}
  (\bibinfo{year}{2000}) \bibinfo{pages}{1403--1410}. \URLprefix
  \url{https://doi.org/10.1016/j.jsv.2020.115423}.
\bibitem[{Noiray and Schuermans(2013)}]{noiray2013deterministic}
\bibinfo{author}{N.~Noiray}, \bibinfo{author}{B.~Schuermans},
\newblock \bibinfo{title}{Deterministic quantities characterizing noise driven
  {H}opf bifurcations in gas turbine combustors},
\newblock \bibinfo{journal}{Int. J. Non-Linear Mech.} \bibinfo{volume}{50}
  (\bibinfo{year}{2013}) \bibinfo{pages}{152--163}. \URLprefix
  \url{https://doi.org/10.1016/j.ijnonlinmec.2012.11.008}.
\bibitem[{Noiray and Denisov(2017)}]{noiray2017method}
\bibinfo{author}{N.~Noiray}, \bibinfo{author}{A.~Denisov},
\newblock \bibinfo{title}{A method to identify thermoacoustic growth rates in
  combustion chambers from dynamic pressure time series},
\newblock \bibinfo{journal}{Proc. Combust. Inst.} \bibinfo{volume}{36}
  (\bibinfo{year}{2017}) \bibinfo{pages}{3843--3850}. \URLprefix
  \url{https://doi.org/10.1016/j.proci.2016.06.092}.
\bibitem[{Bonciolini et~al.(2017)Bonciolini, Ebi, Boujo, and
  Noiray}]{bonciolini2017subcritical}
\bibinfo{author}{G.~Bonciolini}, \bibinfo{author}{D.~Ebi},
  \bibinfo{author}{E.~Boujo}, \bibinfo{author}{N.~Noiray},
\newblock \bibinfo{title}{Subcritical thermoacoustic bifurcation in turbulent
  combustors: effects of inertia},
\newblock in: \bibinfo{booktitle}{26th International Colloquium on the Dynamics
  of Explosions and Reactive Systems (ICDERS 2017)}, \bibinfo{year}{2017}.
\bibitem[{Gopalakrishnan et~al.(2016)Gopalakrishnan, Tony, Sreelekha, and
  Sujith}]{gopalakrishnan2016stochastic}
\bibinfo{author}{E.~Gopalakrishnan}, \bibinfo{author}{J.~Tony},
  \bibinfo{author}{E.~Sreelekha}, \bibinfo{author}{R.~I. Sujith},
\newblock \bibinfo{title}{Stochastic bifurcations in a prototypical
  thermoacoustic system},
\newblock \bibinfo{journal}{Phys. Rev. E} \bibinfo{volume}{94}
  (\bibinfo{year}{2016}) \bibinfo{pages}{022203}. \URLprefix
  \url{http://dx.doi.org/10.1103/PhysRevE.94.022203}.
\bibitem[{Kasthuri et~al.(2019)Kasthuri, Unni, and
  Sujith}]{kasthuri2019bursting}
\bibinfo{author}{P.~Kasthuri}, \bibinfo{author}{V.~R. Unni},
  \bibinfo{author}{R.~I. Sujith},
\newblock \bibinfo{title}{Bursting and mixed mode oscillations during the
  transition to limit cycle oscillations in a matrix burner},
\newblock \bibinfo{journal}{Chaos} \bibinfo{volume}{29} (\bibinfo{year}{2019})
  \bibinfo{pages}{043117}. \URLprefix \url{https://doi.org/10.1063/1.5095401}.
\bibitem[{Nair and Sujith(2015)}]{nair2015reduced}
\bibinfo{author}{V.~Nair}, \bibinfo{author}{R.~I. Sujith},
\newblock \bibinfo{title}{A reduced-order model for the onset of combustion
  instability: physical mechanisms for intermittency and precursors},
\newblock \bibinfo{journal}{Proc. Combust. Inst.} \bibinfo{volume}{35}
  (\bibinfo{year}{2015}) \bibinfo{pages}{3193--3200}. \URLprefix
  \url{https://doi.org/10.1016/j.proci.2014.07.007}.
\bibitem[{Seshadri et~al.(2016)Seshadri, Nair, and
  Sujith}]{seshadri2016reduced}
\bibinfo{author}{A.~Seshadri}, \bibinfo{author}{V.~Nair},
  \bibinfo{author}{R.~I. Sujith},
\newblock \bibinfo{title}{A reduced-order deterministic model describing an
  intermittency route to combustion instability},
\newblock \bibinfo{journal}{Combust. Theory Model.} \bibinfo{volume}{20}
  (\bibinfo{year}{2016}) \bibinfo{pages}{441--456}. \URLprefix
  \url{https://doi.org/10.1080/13647830.2016.1143123}.
\bibitem[{Tandon et~al.(2020)Tandon, Pawar, Banerjee, Varghese, Durairaj, and
  Sujith}]{tandon2020bursting}
\bibinfo{author}{S.~Tandon}, \bibinfo{author}{S.~A. Pawar},
  \bibinfo{author}{S.~Banerjee}, \bibinfo{author}{A.~J. Varghese},
  \bibinfo{author}{P.~Durairaj}, \bibinfo{author}{R.~I. Sujith},
\newblock \bibinfo{title}{Bursting during intermittency route to thermoacoustic
  instability: Effects of slow--fast dynamics},
\newblock \bibinfo{journal}{Chaos} \bibinfo{volume}{30} (\bibinfo{year}{2020})
  \bibinfo{pages}{103112}. \URLprefix \url{https://doi.org/10.1063/5.0005379}.
\bibitem[{Pawar et~al.(2021)Pawar, Raghunathan, Reeja, Midhun, and
  Sujith}]{pawar2021effect}
\bibinfo{author}{S.~A. Pawar}, \bibinfo{author}{M.~Raghunathan},
  \bibinfo{author}{K.~Reeja}, \bibinfo{author}{P.~Midhun},
  \bibinfo{author}{R.~I. Sujith},
\newblock \bibinfo{title}{Effect of preheating of the reactants on the
  transition to thermoacoustic instability in a bluff-body stabilized dump
  combustor},
\newblock \bibinfo{journal}{Proc. Combust. Inst.} \bibinfo{volume}{38}
  (\bibinfo{year}{2021}) \bibinfo{pages}{6193--6201}. \URLprefix
  \url{https://doi.org/10.1016/j.proci.2020.06.370}.
\bibitem[{Nicoud and Wieczorek(2009)}]{nicoud2009zero}
\bibinfo{author}{F.~Nicoud}, \bibinfo{author}{K.~Wieczorek},
\newblock \bibinfo{title}{About the zero {M}ach number assumption in the
  calculation of thermoacoustic instabilities},
\newblock \bibinfo{journal}{Int. J. Spray Combust. Dyn.} \bibinfo{volume}{1}
  (\bibinfo{year}{2009}) \bibinfo{pages}{67--111}. \URLprefix
  \url{https://doi.org/10.1260%2F175682709788083335}.
\bibitem[{McManus et~al.(1993)McManus, Poinsot, and Candel}]{mcmanus1993review}
\bibinfo{author}{K.~McManus}, \bibinfo{author}{T.~Poinsot},
  \bibinfo{author}{S.~M. Candel},
\newblock \bibinfo{title}{A review of active control of combustion
  instabilities},
\newblock \bibinfo{journal}{Prog. Energy Combust. Sci.} \bibinfo{volume}{19}
  (\bibinfo{year}{1993}) \bibinfo{pages}{1--29}. \URLprefix
  \url{https://doi.org/10.1016/0360-1285(93)90020-F}.
\bibitem[{Lieuwen(2021)}]{lieuwen2021unsteady}
\bibinfo{author}{T.~C. Lieuwen}, \bibinfo{title}{Unsteady Combustor Physics},
  \bibinfo{publisher}{Cambridge University Press}, \bibinfo{year}{2021}.
  \URLprefix \url{https://doi.org/10.1017/9781108889001}.
\bibitem[{Lores and Zinn(1973)}]{lores1973nonlinear}
\bibinfo{author}{M.~E. Lores}, \bibinfo{author}{B.~T. Zinn},
\newblock \bibinfo{title}{Nonlinear longitudinal combustion instability in
  rocket motors},
\newblock \bibinfo{journal}{Combust. Sci. and Tech.} \bibinfo{volume}{7}
  (\bibinfo{year}{1973}) \bibinfo{pages}{245--256}. \URLprefix
  \url{https://doi.org/10.1080/00102207308952365}.
\bibitem[{Balasubramanian and Sujith(2008)}]{balasubramanian2008thermoacoustic}
\bibinfo{author}{K.~Balasubramanian}, \bibinfo{author}{R.~I. Sujith},
\newblock \bibinfo{title}{Thermoacoustic instability in a {R}ijke tube:
  Non-normality and nonlinearity},
\newblock \bibinfo{journal}{Phys. Fluids} \bibinfo{volume}{20}
  (\bibinfo{year}{2008}) \bibinfo{pages}{044103}. \URLprefix
  \url{https://doi.org/10.1063/1.2895634}.
\bibitem[{Culick and Kuentzmann(2006)}]{culick2006unsteady}
\bibinfo{author}{F.~Culick}, \bibinfo{author}{P.~Kuentzmann},
  \bibinfo{title}{Unsteady motions in combustion chambers for propulsion
  systems}, \bibinfo{type}{Technical Report}, NATO Research and Technology
  Organization Neuilly-Sur-Seine (France), \bibinfo{year}{2006}.
\bibitem[{Lieuwen(2003)}]{lieuwen2003statistical}
\bibinfo{author}{T.~C. Lieuwen},
\newblock \bibinfo{title}{Statistical characteristics of pressure oscillations
  in a premixed combustor},
\newblock \bibinfo{journal}{J. Sound Vib.} \bibinfo{volume}{260}
  (\bibinfo{year}{2003}) \bibinfo{pages}{3--17}. \URLprefix
  \url{https://doi.org/10.1016/S0022-460X(02)00895-7}.
\bibitem[{Laera et~al.(2017)Laera, Campa, and Camporeale}]{laera2017finite}
\bibinfo{author}{D.~Laera}, \bibinfo{author}{G.~Campa},
  \bibinfo{author}{S.~Camporeale},
\newblock \bibinfo{title}{A finite element method for a weakly nonlinear
  dynamic analysis and bifurcation tracking of thermoacoustic instability in
  longitudinal and annular combustors},
\newblock \bibinfo{journal}{Appl. Energy} \bibinfo{volume}{187}
  (\bibinfo{year}{2017}) \bibinfo{pages}{216--227}. \URLprefix
  \url{https://doi.org/10.1016/j.apenergy.2016.10.124}.
\bibitem[{Minorsky(1962)}]{minorsky1962nonlinear}
\bibinfo{author}{N.~Minorsky}, \bibinfo{title}{Nonlinear Oscillations},
  \bibinfo{publisher}{Princeton, N.J., Van Nostrand}, \bibinfo{year}{1962}.
\bibitem[{Balanov et~al.(2009)Balanov, Janson, Postnov, and
  Sosnovtseva}]{balanov2009simple}
\bibinfo{author}{A.~Balanov}, \bibinfo{author}{N.~Janson},
  \bibinfo{author}{D.~Postnov}, \bibinfo{author}{O.~Sosnovtseva},
  \bibinfo{title}{Synchronization: From Simple to Complex}, Springer Series in
  Synergetics, \bibinfo{publisher}{Springer}, \bibinfo{year}{2009}. \URLprefix
  \url{https://doi.org/10.1007/978-3-540-72128-4}.
\bibitem[{Krylov and Bogoliubov(2016)}]{krylov2016introduction}
\bibinfo{author}{N.~M. Krylov}, \bibinfo{author}{N.~N. Bogoliubov},
  \bibinfo{title}{Introduction to Non-Linear Mechanics.(AM-11)},
  \bibinfo{publisher}{Princeton University Press}, \bibinfo{year}{2016}.
\bibitem[{Stratonovich(1967)}]{stratonovich1967topics}
\bibinfo{author}{R.~L. Stratonovich}, \bibinfo{title}{Topics in the Theory of
  Random Noise}, \bibinfo{publisher}{CRC Press}, \bibinfo{year}{1967}.
\bibitem[{A.~Pavliotis(2014)}]{Grigorios2014stochastic}
\bibinfo{author}{G.~A.~Pavliotis}, \bibinfo{title}{Stochastic Processes and
  Applications: Diffusion Processes, the Fokker-Planck and Langevin Equations},
  \bibinfo{publisher}{Springer}, \bibinfo{year}{2014}. \URLprefix
  \url{https://doi.org/10.1007/978-1-4939-1323-7}.
\bibitem[{Stratonovich(1963)}]{stratonovich1963topics}
\bibinfo{author}{R.~Stratonovich}, \bibinfo{title}{Topics in the Theory of
  Random Noise: General Theory of Random Processes Nonlinear Transformations of
  Signals and Noise}, \bibinfo{publisher}{Gordon \& Breach},
  \bibinfo{year}{1963}.
\bibitem[{Coffey and Kalmykov(2012)}]{coffey2012langevin}
\bibinfo{author}{W.~T. Coffey}, \bibinfo{author}{Y.~P. Kalmykov},
  \bibinfo{title}{The Langevin equation: with Applications to Stochastic
  Problems in Physics, Chemistry and Electrical Engineering},
  \bibinfo{edition}{3} ed., \bibinfo{publisher}{World Scientific},
  \bibinfo{year}{2012}. \URLprefix \url{https://doi.org/10.1142/8195}.

\end{thebibliography}





\end{document}